\documentclass{vldb}
\usepackage{graphicx}
\usepackage{balance}  

\usepackage{booktabs}
\makeatother \makeatletter
\newif\if@restonecol
\makeatother

\usepackage[linesnumbered,ruled,vlined]{algorithm2e}

\usepackage{color}
\usepackage{colortbl}
\usepackage{url}
\usepackage{verbatim}
\usepackage[shortlabels]{enumitem}
\usepackage{amssymb}
\usepackage{multirow}
\usepackage{lscape}
\usepackage{float}
\usepackage[T1]{fontenc}
\usepackage{tgtermes}
\usepackage{listings}

\usepackage{url}

\setlist[itemize]{leftmargin=*}
\setlist[enumerate]{leftmargin=*}

\usepackage{epstopdf}

\definecolor{mygray}{gray}{.9}
\definecolor{mygreen}{rgb}{0.0, 0.5, 0.0}
\definecolor{myred}{rgb}{0.8, 0.0, 0.0}
\definecolor{mygreen1}{rgb}{0.03, 0.91, 0.87}
\definecolor{mypink}{rgb}{1.0, 0.0, 0.5}
\definecolor{mycorn}{rgb}{0.98, 0.93, 0.36}

\newcommand{\yuliang}[1]{{\it\small\textcolor{blue}{[[[ {#1}\ --yuliang ]]]}}}

\newcommand{\wctan}[1]{{\it\small\textcolor{red}{[[[ {#1}\ --wangchiew ]]]}}}

\makeatletter
\def\@copyrightspace{\relax}
\makeatother
\usepackage{dsfont}

\newcommand{\senti}{\mathsf{senti}}

\newcommand{\system}{{\sc OpineDB}}

\vldbTitle{Subjective Databases}
\vldbAuthors{Yuliang Li, Aaron Feng, Jinfeng Li, Saran Mumick, Alon Halevy, Vivian Li, Wang-Chiew Tan}
\vldbDOI{https://doi.org/10.14778/3342263.3342271}
\vldbVolume{12}
\vldbNumber{11}
\vldbYear{2019}

\begin{document} 

\title{Subjective Databases}



\numberofauthors{3} 

\author{
\alignauthor
Yuliang Li 
\alignauthor 
Aaron Feng 
\alignauthor 
Jinfeng Li
\and
\alignauthor
Saran Mumick
\alignauthor  \hspace{-3cm}
Alon Halevy
\alignauthor \hspace{-5cm}
Vivian Li
\alignauthor \hspace{-8cm}
Wang-Chiew Tan\\
\hspace{-19cm}
\affaddr{Megagon Labs} \\
\hspace{-18cm}
\{yuliang, aaron, jinfeng, saran, alon, vivian, wangchiew\}@megagon.ai}

\maketitle

\begin{abstract}
Online users are constantly seeking experiences, such as
a hotel with clean rooms and a lively bar, or a restaurant
for a romantic rendezvous. However, e-commerce search
engines only support queries involving objective attributes
such as location, price, and cuisine, and any experiential data
is relegated to text reviews. 

In order to support experiential queries, a database system
needs to model {\em subjective} data. Users should be
able to pose  queries that specify subjective experiences using their own words, in addition to conditions on the usual objective attributes.
This paper introduces \system, a subjective database system that addresses these challenges. We introduce a data model for subjective databases. We describe how \system\ translates subjective queries against the subjective database schema, which is done by matching the user query phrases to the underlying schema. We also show how the experiential conditions specified by the user can be combined and the results aggregated and ranked.
We demonstrate that subjective databases satisfy user needs more effectively and accurately than alternative techniques through experiments with real data of hotel and restaurant reviews.
\end{abstract}


\section{Introduction}\label{sec:intro}

Database systems model entities in a domain with a set of attributes. Typically, these attributes are objective in the sense that they have an unambiguous value for a given entity, even if the value is unknown to the database, known only probabilistically, or recorded erroneously. Typical examples of such attributes include product specifications, details of a purchase order, or values of sensor readings.  The boolean nature of database query languages reinforces the primacy of objective data--a tuple is either in the answer to the query or is not, but cannot be anywhere in between. 

However, the world also abounds with {\em subjective} attributes for which there is no unambiguous value and are of great interest to users. Examples of such attributes occur in a variety of domains, including the cleanliness of hotel rooms, the difficulty level of an online course, or whether a restaurant is romantic. Currently, the data for these attributes, when it exists, is typically left in text reviews or social media, but not modeled in the database and therefore not queryable. As a result, making decisions that involve subjective preferences is labor intensive for the end user. 
\begin{figure}[!ht]
\centering
\includegraphics[width=0.36\textwidth]{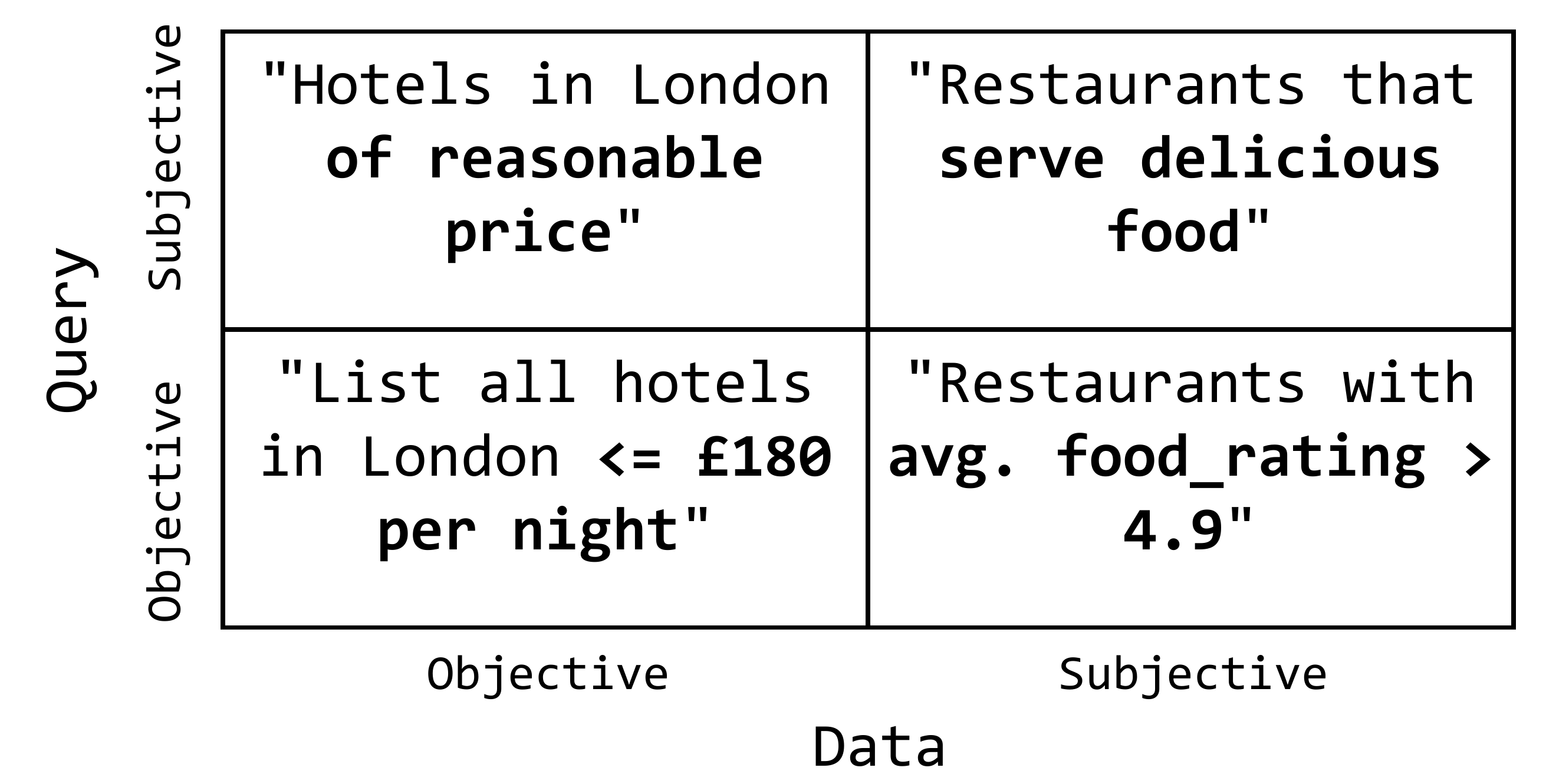}
\caption{\small{Objective/subjective data and queries.}}
\label{fig:quadrants}
\end{figure}

Figure~\ref{fig:quadrants} illustrates that subjectivity can occur in both the data and queries. The lower left quadrant, where both the data and the queries are objective, represents the vast majority of databases today. The lower right quadrant represents how some subjective data has been shoehorned into database systems to date. For example, user ratings of a restaurant are stored in a numerical field, and queries are objective in that they refer to predicates or aggregates on that field (e.g., {\em return restaurants with an aggregate rating of more than 4.5}).  As shown in the upper left hand quadrant, users can pose subjective queries even on objective data. However, in many cases a proper visualization of the answer (e.g., with a histogram or a map) suffices to help the user make their subjective judgment.

This paper introduces the \system\ System, a database system that explicitly models subjective data and answers subjective queries, thereby addressing the challenges in the upper right quadrant of Figure~\ref{fig:quadrants}. With this capability, \system\  enables a new set of applications where users can search by their subjective preferences. 
 
\subsection{Example: experiential search}
An important motivating application for subjective databases is {\em experiential search}. Today's e-commerce search engines support querying by objective attributes of a service or product, such as price, location or square footage. However,  as we illustrate in Section~\ref{sec:subjective-queries}, users overwhelmingly want to be able to search by specifying the experiences they desire, and these  are often expressed as subjective predicates. Table~\ref{fig:queries} shows examples of queries that users should be able to pose. 

\renewcommand{\arraystretch}{1.3}
\begin{table}[!ht]
\small
\caption{\small{Queries that express experiential preferences.}}
\label{fig:queries}
\begin{tabular}{  p{0.11\textwidth} | p {0.32\textwidth} }
\hline
 {\bf Domain} & {\bf Example Query}  \\ \hline
Hotels &  {\em a hotel with a lively}   {\em bar scene and clean rooms} \\ \hline
Dining &  {\em a restaurant with a sunset}  {\em view of Tokyo Tower} \\ \hline
Employment & {\em a job with a dynamic team} {\em working on social good}  \\ \hline
Housing &  {\em a 2-bedroom apartment in a quiet}  {\em neighborhood near good cafes}  \\ \hline
Online education & {\em a 1-week course on python} {\em with short programming exercises} \\ \hline
Travel & {\em a relaxing trip to a beach}  {\em on the Mediterranean}  \\ \hline
Multi-domain & {\em a quiet Thai restaurant next}  {\em to a cinema that shows Ocean's 8}  \\ \hline
\end{tabular}
\vspace{-5mm}
\end{table}
\renewcommand{\arraystretch}{1.0}

We use the domain of hotel search to illustrate some of the challenges of experiential search. A subjective database in the hotel domain will have a schema that models hotel rooms with a mixture of objective and subjective attributes. 

Consider a user who is searching for a hotel in London that costs less than 180 pounds per night,  has really clean rooms, and is a romantic getaway. The first condition is objective and simple to satisfy, but the second and third conditions are subjective. Conceivably, techniques from sentiment analysis and opinion mining~\cite{liu2012sentiment} can be used to extract relevant descriptions from hotel reviews. However, \system\ faces several novel challenges. 
 
 First, \system\ needs to aggregate the reviews in a meaningful way and so they can be queried effectively and efficiently. To do so, \system\ introduces the concept of {\em markers}, which are the distinctions about the domain that the designer thinks are important for the application. \system\ then aggregates phrases from the reviews along these markers to create {\em marker summaries}.  An application based on \system\ may also decide to expose these markers in the user interface for easier querying. 
 The choice of markers is based on a combination of mining the review data and knowing the requirements of the  application and is crucial to the quality of the query results.  For example, the designer might decide that room cleanliness can be modeled by $\{$clean, dirty$\}$, while bathroom style requires a  scale such as $\{$old, standard, modern, luxurious$\}$. 
 
 Second, \system\ needs to  answer complex queries in a principled way. In our example, \system\ has to combine two subjective query predicates and an objective predicate. The user may also decide to complicate the query and only consider opinions of people who reviewed at least 10 hotels. In that case, \system\ needs to refer back to the review data and consider only reviews by qualified reviewers, which will involve recalculating the room cleanliness scores for every hotel. 
 
 Finally, users may not always use terms that fit neatly into the database schema. For example, the user may ask for romantic hotels, but the schema does not have an attribute for romantic. However, \system\ may have background knowledge that suggests that hotels with exceptional service
 and luxurious bathrooms are often considered romantic. Since the quality of service and bathroom luxury are captured in its schema as subjective attributes, \system\ can reformulate the query for romantic rooms into a combination of attributes in the schema. 
Users may also query for properties that are not even close to the database schema, such as hotels that are good for motorcyclists. In this case, \system\ will verify this requirement by falling back on text search in the reviews to see if any reviews mention amenities for motorcyclists. $\quad\square$ 
\smallskip

The above example illustrates the broader dichotomy that exists between the search for structured objects and search for documents. As a typical example,  online shoppers will typically go to a search engine like Google or Bing to find the top-rated espresso machines, and then go to Amazon to purchase one (a situation that none of these companies are happy with and are working hard to tilt in their favor). The fundamental reason for the dichotomy is that  the experiential aspects of the object (or service) being purchased are not {\em queryable}, and therefore users rely on Web search engines as the best option to discover them. This paper focuses on the core issues involved in making subjective data queryable, but ultimately these techniques can be used to extend database systems as well as document retrieval systems.

\subsection{Contributions}

Specifically, we make the following contributions:
\begin{itemize}\parskip=1pt
\item We introduce a data model for subjective databases and an associated variant of SQL that supports subjective predicates for querying such databases.
The key feature of a subjective database is that every
subjective attribute is associated with a new data type called the \emph{linguistic domain}, which 
is a set of phrases for describing the attribute. 
The designer specifies a set of {\em markers} that are the phrases in the linguistic domain 
that represent the important concepts of the domain.
The linguistic domain and markers are effective intermediaries between text and queries;
They help produce meaningful aggregation from information extracted from text and help produce high-quality answers to queries.
\item We present \system, our query processing system for subjective databases. \system\ 
effectively interprets subjective predicates against the subjective database schema through a combination of NLP and IR techniques; by matching against the linguistic domains and markers, or finding correlations between subjective predicates and attributes in reviews. It is also able to fall back to exploiting traditional 
text retrieval methods as needed.
After interpretation, \system\ uses a variant of fuzzy logic to combine the results of multiple subjective query predicates. 
\item One of the major challenges that \system\ raises is the construction of the subjective database, that requires extracting relevant information from text and designing the subjective database schema.
We developed a novel extraction pipeline which requires little NLP expertise from the schema designer 
and also facilitates the automatic discovery of potential markers for subjective attributes. 
We show that our extraction pipeline achieves state-of-the-art performance by leveraging the most recent advances in NLP techniques such as BERT~\cite{bert}.
\item We demonstrate the effectiveness and efficiency of \system\ with real-world review data from two domains --- hotels and restaurants.
Our experimental results demonstrate the need for subjective databases, 
\system\ outperforms two search baselines by up to 15\% and 10\% even when evaluated conservatively, 
and \system\ achieves a speedup of up to 6.6x through the use of marker summaries for query processing. 
\end{itemize}

\noindent
{\bf Outline:~}
We introduce our data model for subjective databases and  
subjective queries in Section \ref{sec:datamodel}. 
Section~\ref{sec:life} describes query processing in \system. 
Section \ref{sec:aggregation} describes how \system\
constructs a subjective database by extracting opinions from reviews and aggregating them
into summaries. We present our experimental results in Section \ref{sec:experiments}.
We discuss related work in Section \ref{sec:related}.
We also provide more details and examples in the appendix of the full version \cite{opinedb_arxiv}.

\newcommand{\oa}{\mathsf{OA}}
\newcommand{\sa}{\mathsf{SA}}
\newcommand{\dom}{\mathsf{dom}}
\newcommand{\od}{\mathsf{OD}}
\newcommand{\sd}{\mathsf{SD}}
\newcommand{\st}{\mathsf{ST}}
\newcommand{\ex}{\mathsf{EX}}
\newcommand{\et}{\mathsf{ET}}
\newcommand{\at}{\mathsf{AT}}
\newcommand{\op}{\mathsf{OP}}
\renewcommand{\sp}{\mathsf{SP}}
\newcommand{\aggr}{\mathsf{aggr}}

\section{Data Model}
\label{sec:datamodel}

A relation in \system\ includes objective and subjective attributes. Informally, a subjective attribute represents an aggregate view of textual phrases extracted from reviews. 
In this section we introduce {\em linguistic domains} that capture these phrases and {\em marker summaries} that represent the aggregates.

Consider the example of an attribute {\tt room\_cleanliness} in the domain of hotels. The raw data for this attribute exists in reviews and social media and consists of a wide variety of phrases such as 
\begin{enumerate}\parskip=0pt
    \item ``{\em The floor in my room was filthy dirty ...}'',
    \vspace{-1mm}
    \item ``{\em The room was clean, well-decorated and ...}'', or
      \vspace{-1mm}
    \item ``{\em Spotlessly clean and good location'}'.
\end{enumerate}

The challenge \system\ faces is to {\em aggregate} these phrases into a meaningful signal and to rank hotels appropriately in response to queries that may themselves include different linguistic phrases. The ability to aggregate linguistic phrases is one of the key aspects that distinguishes \system\ from text retrieval systems where ranking is typically based on similarity of unstructured text.

To define an aggregation function, there needs to be a scale onto which the aggregation is performed. However, when dealing with text, we cannot always arrange the main landmarks in the domain into a linearly-ordered scale.  Hence, \system\ lets the schema designer define a set of {\em markers} in the domain that represent the points onto which we map the reviews. In our example, the markers might be a linear scale 
$[$very\_clean, average, dirty, very\_dirty$]$ 
for 
room cleanliness and a set of markers for bathroom styles
may be [old, standard, modern, luxurious], which
is a set of categories (not a linear scale) that
describes the different styles of bathroom.
 \system\ maintains a {\em marker summary}, which is a view that aggregates the phrases from the reviews onto the markers.
  Specifically,  \system\  computes a value representing the membership of each hotel for each marker. For example, the record
 $[$very\_clean: 20, average: 70, dirty: 30, very\_dirty: 10$]$ for a hotel would represent that the hotel is closer to being a member of {\tt average} than to the other markers. As we see later, there  are different possible aggregation functions \system\ can use and the appropriate choice depends on the semantics of the attribute. 
 At present, \system's marker summaries are histograms that tabulate the number of phrases of reviews closest to each marker.
 Each marker summary also records features useful for query processing including the average sentiment score and the average phrase embedding vector.
 
 In addition to aggregating the raw data, the marker summaries serve two other important goals. 
 First, by aggregating the review data offline, query processing can be much more efficient. 
 Accessing the raw data at query time would be prohibitively expensive. In our experiments,
 query processing was accelerated by a factor of up
 to  6.6x by only accessing the marker summaries. 
 Second, the markers enable the system designer to shape which important aspects of the reviews to highlight in the application.  Even though \system\ allows 
 end users to specify arbitrary keywords in their query, 
 markers can be useful for end users and applications
 in gauging the range of linguistic terms that are supported by marker summaries.

 We now describe each of the above components in detail.

\smallskip
\noindent
{\bf Linguistic domains:} A {\em linguistic domain} is defined to be a set of short linguistic phrases 
(which we refer to as {\em linguistic variations}) that describe a particular aspect of an object. 
The linguistic domain allows \system\ to capture the different ways 
of describing the cleanliness of a room.
For example, \{ ``very clean'', ``pretty clean'', ``spotless'', ``average'', ``not dirty'', ``dirty'', ``stained carpet'', ``very dirty''...~\}
is a linguistic domain for the cleanliness aspect of the room object. 

The linguistic domain is not enumerated in advance. \system\ bootstraps it by extracting phrases from the review data. 
Phrases in reviews can express  opinions about an object directly (e.g., ``very clean room''), or 
indirectly (e.g., ``the room has stained carpet''). 
\system's extraction module supports both types of opinions.

\smallskip
\noindent
{\bf Markers and Marker summaries:} A {\em marker summary} is defined over a linguistic domain and is a record type 
$\mbox{Rcd}:[f_1, ..., f_k]$ 
where \mbox{Rcd} is the name of the record,
$f_i$ are names of {\em markers}, and the type for each field is assumed to be a number. There are two types of marker summaries: {\em linearly-ordered} or {\em categorical}. A linearly-ordered marker summary is one where $f_1, ..., f_k$ form a linear scale.  An example of a such a marker summary
for room cleanliness over the linguistic domain described earlier is shown below.

\vspace{-1mm}
\begin{center}
room\_cleanliness : [very\_clean, average, dirty, very\_dirty]   
\end{center}
\vspace{-1mm}

\noindent
A phrase can contribute {\em in part} to multiple markers of a linearly-ordered marker summary. For example, the phrase ``rooms are quite clean'' can contribute in equal proportions (0.5 each) 
to the markers ``very\_clean'' and ``average''. 
An example instance of the \\ room\_cleanliness marker summary is 
$[$very\_clean: 90.5, average: 60.5, dirty: 30, very\_dirty: 20$]$. 
We use the term ``marker summary'' to refer to either the record type or the record instance when the context is clear.

Note that phrases in a linguistic domain do not always fall into a simple linearly-ordered scale of sentiment. For example, the fields we obtained by mining reviews from \url{booking.com} show that the quietness of a room may be described with words such as ``annoying'', ``peaceful'', ``very noisy'', ``traffic noise'', ``constant noise'', which do not follow a natural linear order.
To handle such cases, we also allow for categorical marker summaries in \system.
A categorical marker summary is one where no two markers form a linear
scale. An example of a categorical marker summary is 
\begin{center}
style : [old, standard, modern, luxurious]   
\end{center}
\noindent
A phrase can contribute {\em as a whole} to multiple markers in a categorical marker summary. For example,  ``extravagant old-fashioned bathrooms'' contributes
to both ``old'' and ``luxurious'' (1 count each). 

At present, we assume that a marker summary is either linearly-ordered or categorical.
Of course, for each categorical marker such as ``luxurious'', there may be
a linearly-ordered marker summary on the degree of ``luxuriousness''. As with any database design, it is the task of the schema designer to decide the appropriate  level of granularity to model the domain.
In our context, \system\ assists 
  her by clustering the linguistic domain (See Section~\ref{sec:domainmarkers} for details).

\smallskip
\noindent
{\bf Schema of a subjective database:} The schema of an \system\ application consists of three elements: (1)
the main schema that is visible to the user and the application programmer,
(2) the raw review data, and (3) the extractions of relevant phrases from the reviews from which we compute marker summaries. Parts (2) and (3) of the schema are intended to support queries that might qualify the reviews (e.g., consider only reviewers who have reviewed more than~10 hotels) and to support \system's ability to fall back on raw text when a query cannot be answered using the database schema. We discuss some of the details of these components of the schema in Section~\ref{sec:aggregation}. In what follows we focus on (1), which illustrates the main novel aspects of our data model. 

\begin{figure}[t]
\begin{minipage}{0.55\textwidth}
\small
\begin{tabbing}
{\bf Hotels} 
\begin{tabular}{|l|l|l|l|}
\hline
hotelname &  capacity &  address & price\_pn \\
\hline
\end{tabular}
\end{tabbing}
\normalsize

\small
\begin{tabbing}
{\bf HRoomCleanliness} \=
\begin{tabular}{|l|l|}
\hline
hotelname & $\ast$ {\em room\_cleanliness} \\
\hline
\end{tabular} \\

{\bf HBathroom} \>
\begin{tabular}{|l|l|}
\hline
hotelname & $\ast$ {\em style} \\
\hline
\end{tabular} \\

{\bf HService} \>
\begin{tabular}{|l|l|}
\hline
hotelname & $\ast$ {\em service} \\
\hline
\end{tabular} \\

{\bf HBed} \>
\begin{tabular}{|l|l|}
\hline
hotelname & $\ast$ {\em comfort} \\
\hline
\end{tabular}
\end{tabbing}
\normalsize
\vspace{-1mm}
\noindent
\flushleft
\small
\vspace{-2mm}
\underline{\bf Marker Summaries}\\
\vspace{-2mm}
\begin{tabbing}
$\ast$ room\_cleanliness: \= [very\_clean, average, dirty, very\_dirty] \\
$\ast$ style: \> [old, standard, modern, luxurious] \\
$\ast$ service: \> [exceptional, good, average, bad, very\_bad] \\
$\ast$ comfort: \> [very\_soft, soft, firm, very\_firm, ok, worn\_out]
\end{tabbing}
\normalsize
\end{minipage}
\vspace{-2mm}
\caption{\small A subjective database schema for the hotel domain. Subjective attributes are prefixed with $\ast$.}
\label{fig:subjdbschema}
\vspace{-5mm}
\end{figure}
\normalsize


The schema that is visible to the user or the application is a finite sequence of relation schemas each
of the following form: $R$($K$, $A_1$, $\ldots$, $A_n$)
where $K$ is the key for $R$ (for simplicity, we assume it's a single attribute),  and  $A_1, \ldots, A_m$
are a set of attributes.

We distinguish between two types of attributes. 
An {\em objective attribute} is an attribute whose value 
is based on facts and is largely indisputable.
In contrast, there is no ground 
truth for the value of a {\em subjective attribute}.
The value of a subjective attribute is ``influenced by or 
based on personal beliefs or feelings, rather than based on 
facts''.\footnote{\small{\url{https://dictionary.cambridge.org/us/dictionary/english/subjective}}}
Figure~\ref{fig:subjdbschema} shows an example of a subjective database schema for the hotel domain.

The type of a subjective attribute is a marker summary over a linguistic domain.
The linguistic domain of the subjective attribute {\tt style}, for example,  is a set of phrases that may include \{ ``modern faucets'', ``old shower'', ``OK'', ``adequate'', ``luxurious bath towels'', ... \}. 
As described earlier, the intuition is that the marker summary keeps a summary (e.g., histogram) of 
the subjective phrases for that attribute w.r.t.\ the markers.


The key of the Hotels relation is {\tt hotelname} and it has three objective attributes, {\tt capacity}, {\tt address},
and {\tt price\_pn}. There are four additional relations with the same key attribute
 that contain subjective attributes. 
The attributes {\tt room\_\-cleanliness}, {\tt style}, {\tt service}, {\tt comfort}
are subjective 
attributes of the relations HRoomCleanliness, HBathroom, HService, 
and HBed
respectively and their marker summaries are shown at the bottom of Figure~\ref{fig:subjdbschema}. 


A core issue that \system\ needs to address is that of aggregating a large collection of linguistic phrases onto marker summaries, which we will describe in Section~\ref{sec:aggregation}. In what follows, we describe the query language
of \system.

\vspace{1mm}
\noindent
{\bf Query:~} The \system\ query language is essentially 
SQL with the extra ability of specifying atomic conditions in natural language in the {\tt where} clause. 
For the purposes of this paper, 
we assume that an \system\ query consists of a single {\sf select}-{\sf from}-{\sf where} clause. 
For example, assume that our hotel database is for hotels in London, then
the query for hotels in London that cost less than 150 Euros per night, has clean rooms, and is good as a romantic getaway
can be expressed as shown below:

\renewcommand{\arraystretch}{1.1}
\vspace{1mm}
\begin{tabular}{p{.8cm}p{3in}}
{\sf select} &  * \hspace{0.3cm}{\sf from} {\tt \ Hotels}\\
{\sf where}  & {\tt price\_pn} $<$ 150 {\sf and}  \\
             & ``{\em has really clean rooms}'' {\sf and} ``{\em is a romantic getaway}''
\end{tabular}
\vspace{1mm}
\renewcommand{\arraystretch}{1.0}

The {\tt where} clause is a 
logical expression over a set of conditions.
The example above shows a conjunction of a condition on price and two 
{\em query predicates} (or 
{\em predicates} in short),
which are conditions specified in natural language and they need not be phrases that occur in the reviews.
The user can also directly query for clean rooms using the attribute {\tt room\_cleanliness} in the HRoomCleanliness relation. 
However, to do so she will need to understand the exact semantics of the schema and specify the precise predicate, such as whether  ``clean rooms'' should be interpreted as ``very clean'', ``average'', ``dirty'' or ``very dirty'' rooms. 
By extending the query language to accept natural language predicates, we can support a broader range of user interfaces to subjective databases. We will describe how
\system\  automatically compiles the query predicates against the underlying schema.

Of course, natural language queries can involve non-atomic conditions, but \system\ relies on techniques such as~\cite{zhong2017seq2sql} to decompose a complex query into atomic conditions. As such, we assume atomic query predicates throughout this paper.


As we shall describe in Section~\ref{sec:life}, our subjective query interpreter compiles 
the query predicate ``{\em has really clean rooms}'' into a predicate over the {\tt room\_cleanliness} attribute and
the query predicate ``{\em is a romantic getaway}'' into a predicate over the {\tt service} and {\tt bathroom} attributes. 
If the query predicate cannot be satisfactorily interpreted into a predicate over the existing schema, \system\ falls back to the source reviews to arrive at a ranked set of answers.

\begin{figure}[!ht]
    \centering
    \includegraphics[width=0.46\textwidth]{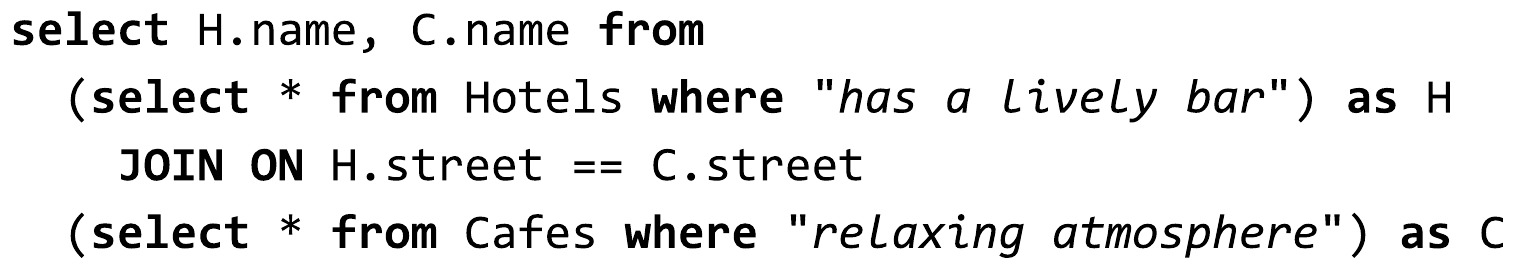}
    \vspace{-3mm}
    \caption{\small{An example of an \system\ query with joins.}}
    \vspace{-2mm}
    \label{fig:sql_example}
\end{figure}

\smallskip
\noindent
\textbf{Benefits of a query language:} One of the 
major advantages of \system\ is that subjective data can be queried declaratively, and therefore we can express complex queries. For example, the semantics of the following query is well defined: 
\emph{``find hotels with clean room based on reviews after 2010''}. Another important example is queries involving joins, such as 
\emph{``find a hotel with a lively bar on the same street as a cafe with a relaxing atmosphere.''} 
(Figure \ref{fig:sql_example}, we leave the discussion of the join semantics to future work).
Being implemented on an RDBMS, \system\ is also able to leverage any query optimization capability of
the underlying engine to boost the querying performance. 
\vspace{-2mm}

Next, we describe how \system\ processes queries.
Although we continue to exemplify our technical discussions with examples from the hotel domain, the techniques we develop are not dependent on a particular domain. In fact, we have conducted our experiments in Section~\ref{sec:experiments} on two domains: 
hotels and restaurants.

\newcommand{\mf}{\mathsf{mf}}
\newcommand{\rep}{\mathsf{rep}}
\newcommand{\similarity}{\mathsf{similarity}}
\newcommand{\avg}{\mathsf{AVG}}
\newcommand{\cosine}{\mathsf{cos}}
\newcommand{\sigmoid}{\mathsf{sigmoid}}
\newcommand{\subp}{\mathsf{subp}}
\newcommand{\idf}{\mathsf{idf}}
\newcommand{\wtv}{\mathsf{w2v}}
\newcommand{\freq}{\mathsf{freq}}
\newcommand{\ta}{$\mathsf{TA}$}

\section{Processing Subjective Queries}
\label{sec:life}


\begin{figure}[t]
    \centering
    \includegraphics[width=.47\textwidth]{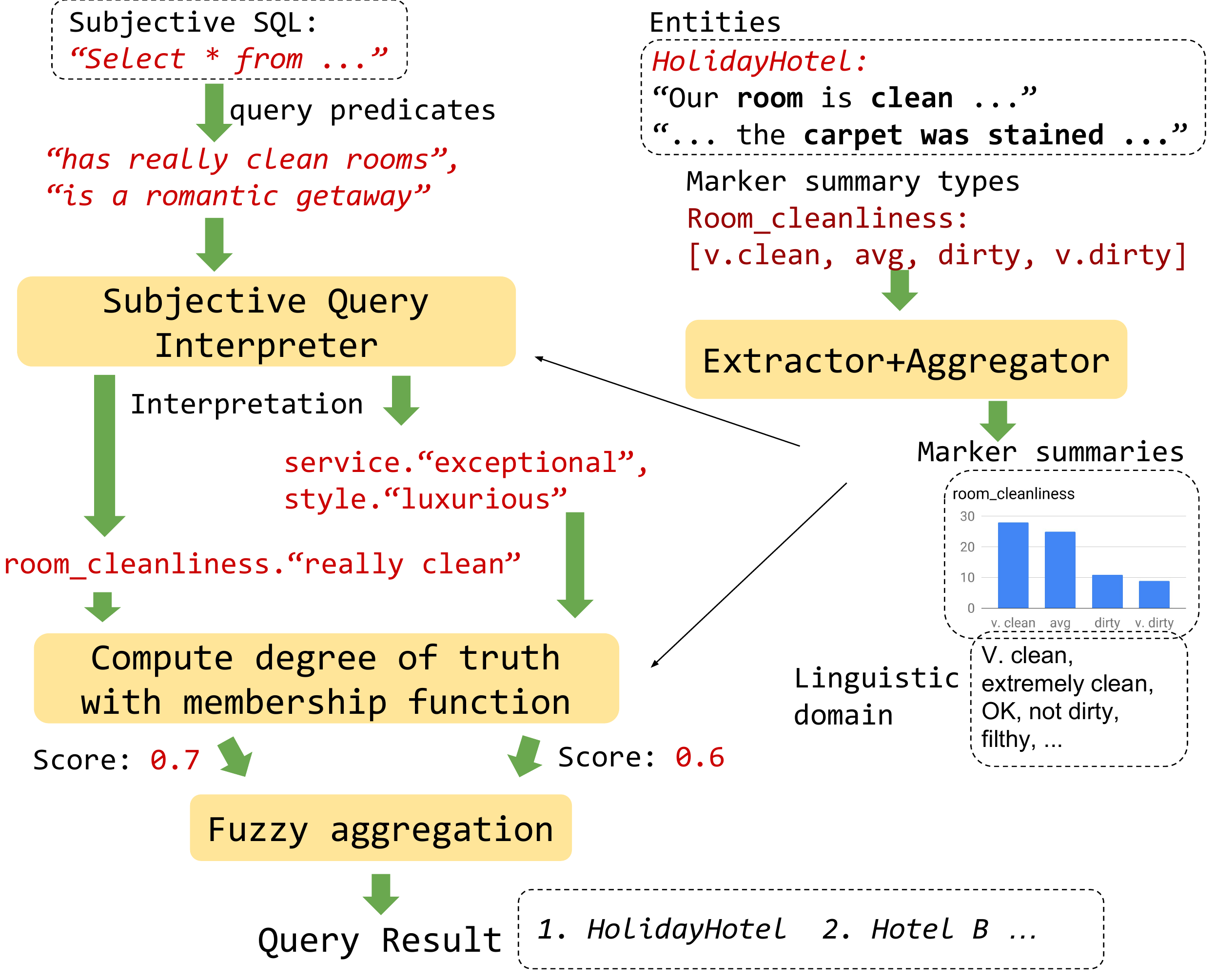}
    \vspace{-4mm}
    \caption{\small{Processing subjective queries with \system.}}
    \vspace{-6mm}
    \label{fig:architecture}
\end{figure}

 To highlight the technical challenges \system\ faces in query processing, we begin with a simple class of queries and then move on to more complex ones.
Figure~\ref{fig:architecture} illustrates the entire process.

\subsection{Predicates with markers}
\label{sec:simple-predicates}

We begin with queries that contain predicates where each predicate maps to a specific subjective attribute and one of its markers. In this discussion we ignore objective attributes because they do not introduce new challenges.

Consider a subjective query that contains a conjunction
of the following 
query predicates:

\renewcommand{\arraystretch}{1.1}
\smallskip
\begin{tabular}{p{1cm}p{3in}}
{\sf select} &  * \hspace{0.3cm}{\sf from} {\tt Hotels} $h$\\
{\sf where}  & ``{\em has firm beds}'' and ``{\em has luxurious bathrooms}''
\smallskip
\end{tabular}
\renewcommand{\arraystretch}{1.0}

Here, we will assume that the query predicates can be interpreted into the following subjective attributes and their respective markers: {\tt comfort}.``firm'' and {\tt style}.``luxurious''. In general, however, a query predicate might not match exactly with one specific marker of a subjective attribute.
Mapping the predicate approximately and into multiple subjective attributes are necessary in many cases.
Section~\ref{sec:predicateinterpretation} describes in more detail how \system\ interprets arbitrary query predicates.

Once we have the interpretation of each query predicate, we need to compute the {\em degree of truth} for each
interpreted predicate for each entity in the database.
We describe that process in detail in Section~\ref{sec:dot}. In this simple case, since the query references markers of the subjective attribute, we assume that the degrees of truth have been computed in advance. Hence, we can focus on the last part of query processing which is to {\em combine} the degrees of truth of multiple predicates in a principled fashion.


\smallskip
\noindent
{\bf Combining degrees of truth  }
The degrees of truths are combined using fuzzy logic~\cite{klir1995fuzzy,fagin1996combining}.
Fuzzy logic generalizes propositional logic by allowing
truth values to be real numbers in the range $[0, 1]$. 
The real truth value of a logical formula $\varphi$
represents the degree of $\varphi$ being satisfied,
where a higher value means a higher degree of satisfying $\varphi$.

With our previous example, we now have the following
query.  The logical AND 
is replaced with $\otimes$ and 
the query predicates 
have been replaced with their respective
interpretations.

\renewcommand{\arraystretch}{1.1}
\smallskip
\begin{tabular}{p{1cm}p{3in}}
{\sf select} &  * \hspace{0.3cm}{\sf from} {\tt Hotels} $h$\\
{\sf where}  & $h$.comfort $\doteq$ ``firm'' $\otimes$ $h$.style $\doteq$ ``luxurious''
\end{tabular}
\vspace{1mm}
\renewcommand{\arraystretch}{1.0}

In the query above, $h$.comfort denotes the marker summary of the bed comfort of hotel $h$. 
The condition $h$.comfort $\doteq$ ``firm'' computes
the degree of truth
of how well the summary of bed comfort of hotel $h$
represents the word ``firm''. Similarly, a degree of truth
is computed for $h$.style $\doteq$ ``luxurious''.

The fuzzy logical operator AND is denoted with $\otimes$.
Later we will see queries with OR, which will be denoted by
 $\oplus$.
In the most classic variant of fuzzy logic \cite{fagin1996combining}, 
$x \otimes y$ is interpreted as MIN$(x,y)$, NOT$(x)$ is interpreted as $(1-x)$
and $x \oplus y$ is interpreted as \\ MAX$(x,y)$. 
Other variants include the multiplication variant~\cite{tnorm} which we use in \system. 
In this variant, following De Morgan's law,  $x \otimes y$ is simply $xy$, negation is $1-x$ and $x \oplus y$ is 
$1-(1-x)*(1-y)$.
Note that an objective predicate will simply be interpreted as 0 or 1. 

Going back to our example, the conditions in the {\sf where} clause are combined using the multiplication variant, which takes the product of the two degrees of truth. 
The result is a ranked list of hotels based on the final degree of truth that the entire  expression in the {\tt where} clause evaluates to.


\smallskip
\noindent
{\bf Why fuzzy logic?}
An alternative to fuzzy logic would be to translate a subjective SQL query 
into a classic SQL query with hard selection constraints. 
For example, the previous two conditions can be written as: 

\vspace{2mm}
\noindent
($h$.{\tt comfort} $\doteq$ ``firm'') $>$ 0.8   and
($h$.{\tt style} $\doteq$ ``luxurious'') $>$ 0.6 
\vspace{2mm}

\noindent
where the thresholds 0.8 and 0.6 are specified
by the application or the end-user. Aside from the inherent difficulty of specifying such thresholds in a meaningful fashion, this method 
may miss entities that
may fall slightly out of the specified constraints.
For example, a hotel with {\tt comfort} score slightly less than 0.8
will be discarded from the result set for
the above query. In contrast, the fuzzy interpretation is more forgiving and may therefore yield entities 
with good overall relevance to the query even if it may
not satisfy the threshold of 0.8 on comfort. (See Appendix A of \cite{opinedb_arxiv} for a visual illustration of this point). 
Furthermore, as the number of conditions increases, 
the number of relevant entities that are potentially missed by the hard constraints only increases. 

\subsection{Predicates with arbitrary phrases}
\label{sec:predicateinterpretation}

In the previous section, the query predicates were simple in the sense that it was clear which subjective attribute they refer to and which value (the marker) the user is specifying. The designer of an \system\ application may constrain the user to such queries, but one of the important benefits of subjectivity is that users can specify queries using their own terms. This, in turn, raises two challenges:
\begin{itemize}\parskip=0pt
\item Interpreting the phrase specified by the user: The user may specify a phrase for a subjective attribute that is not a marker. For example, she may ask for a hotel with rooms that are ``{\em really clean}'' or ``{\em meticulously clean}''. 
In some cases, these phrases may be in the 
linguistic domain of the subjective attribute and in other cases it may be a phrases that the application has never seen before.
\item Determining the subjective attribute(s): in the simplest case, the challenge is to map the predicate to a single subjective attribute (e.g., mapping the predicate {\em ``has really clean rooms''} to the subjective attribute \texttt{room\_cleanliness} with the phrase \emph{``really \\ clean''}). However, the user may specify predicates that do not correspond directly to a subjective attribute, such as ``{\em is a romantic getaway}''. In this case, a combination of subjective attributes may be equivalent to the predicate, or may at least provide strong evidence for it. For example, \system\ may know that hotels with {\tt service}.{\em ``exceptional''} and \texttt{bathroom}.\emph{``luxurious''} are usually considered romantic. In other cases, the user may specify a predicate that does not correspond to any subjective attribute such as ``{\em has great towel art}''.
\end{itemize}
The subjective query interpreter is the component of \system\ that translates 
query predicates onto subjective attributes and their markers or to combinations thereof.  
The interpreter computes an  \emph{interpretation} for every predicate in the query.  
The interpretation consists of expressions of the form  $A.m$ 
where $A$ is an subjective attribute and $m$ is a marker of $A$. 
The goal of the interpreter is to find the expression over the $A.m$'s that best matches $q$. 
Each $A.m$ replaces the original query predicate
as a condition $A \doteq m$ (e.g., {\tt comfort} $\doteq$ ``firm'').
If a predicate interprets into multiple $A.m$'s, \system\ replaces 
the original query predicate as a disjunction of the results. 
For example,
there are two subjective query predicates in our running example: ``{\em has really clean rooms}'' and 
``{\em is a romantic getaway}''. The predicate
``{\em has really clean rooms}'' is
interpreted into \texttt{room\_cleanliness}.\emph{``very clean''} 
due to the high similarity between the predicate with the marker \emph{``very clean''}.
However, the second predicate does not bear sufficiently high similarity to any of the existing markers. In this case, \system\ 
uses an alternative approach to map
the predicate to markers by finding
all markers that frequently co-occurs
with the predicate. 
With this approach, the second predicate is interpreted into 
a disjunction of {\tt service}.{\em ``exceptional''}
and \texttt{bathroom}.\emph{``luxurious''} because the phrase ``romantic getaway'' frequently co-occurs
with exceptional service or luxurious bathroom in the review corpus. Note that ``exceptional service'' or ``luxurious bathrooms'' may not reflect the true meaning of ``romantic getaway''. 
However, they are proxies of ``romantic getaway'' derived in an entirely data-driven way based on the reviews.
We obtain the following fuzzy SQL snippet after this step.

\renewcommand{\arraystretch}{1.1}
\vspace{2mm}
\begin{tabular}{p{1cm}p{3in}}
{\sf select} &  * \hspace{0.3cm}{\sf from} {\tt Hotels} $h$\\
{\sf where}  & {\tt price\_pn} $<$ 150 $\otimes$ \\
             & $h$.room\_cleanliness $\doteq$ ``really clean'' $\otimes$\\
             & ($h$.service $\doteq$ ``exceptional'' $\oplus$  $h$.style $\doteq$ ``luxurious'')
\end{tabular}
\vspace{1mm}
\renewcommand{\arraystretch}{1.0}

As we mention later, the co-occurrence method sometimes
outputs a conjunction of $A.m$'s instead. For example, if 
``exceptional service'' and ``luxurious bathrooms'' are frequently mentioned together along with romantic getaway 
instead of individually, then the above $\oplus$ will be $\otimes$ instead.

As noted earlier,  it may not be possible to completely interpret a query predicate in terms of the database schema. 
Hence, in parallel with trying to interpret the query, \system\ also relies on a text retrieval system 
(described later in this section) to produce matching scores between database entities and query predicates. 
In principle, \system\ should combine the scores of the interpreter and the text retrieval system to 
produce the final ranking. In our current implementation, \system\ uses a threshold to determine whether 
it has enough confidence in the interpretation, and only if it does not, \system\ falls back on the text-retrieval method. 

\begin{figure}[t]
    \centering
    \includegraphics[width=0.39\textwidth]{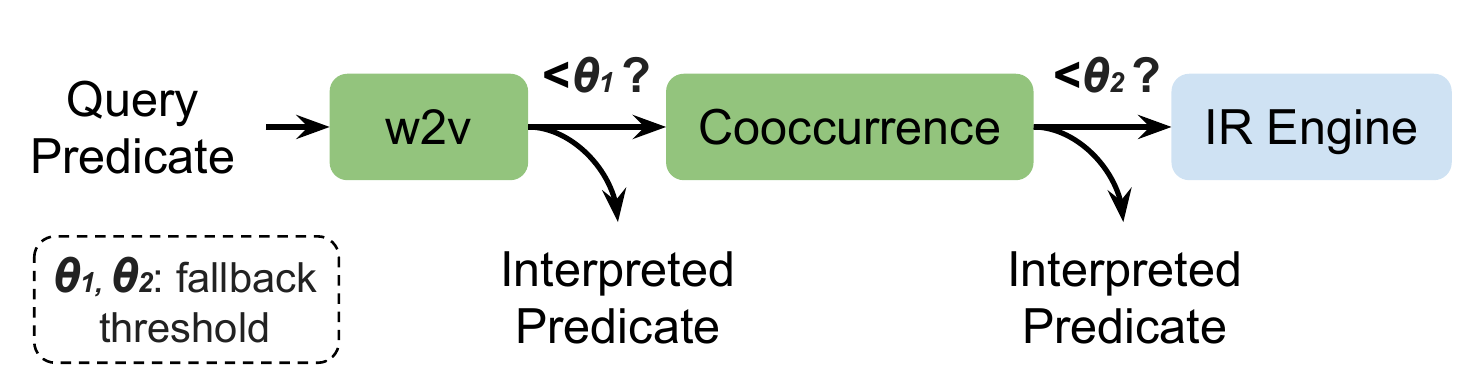}
    \vspace*{-4mm}
    \caption{\small Fallback mechanisms in \system. }
    \label{fig:fallback}
    \vspace{-6mm}
\end{figure}

\smallskip
\noindent
{\bf Predicate interpretation algorithm~}
This algorithm takes as input a query predicate
and returns as output an expression over the set of $A.m$'s as introduced above. 
\system\ currently uses a three-stage approach to interpret query predicates in a best-effort manner (see Figure~\ref{fig:fallback}):
it first applies the word2vec method to find a direct interpretation of the query predicate. If this method fails to produce a satisfactory interpretation, it uses the co-occurrence method to find an approximate interpretation of the query predicate. If the second method fails to produce a satisfactory interpretation, it falls back to the text retrieval method to produce a ranked list of answers.
When an interpretation is successfully obtained from the first or second method, the SQL query is rewritten based on the interpretation and executed to obtain
a ranked list of result. 

\smallskip
{\bf Word2Vec method:~}
Given a query predicate, this method finds the linguistic variations of all subjective attributes
having the highest similarity with the query predicate and returns the attributes and markers
that correspond to the most similar variations as the interpretation.
This method is based on the observation that most query predicates are simple so 
they likely to match closely with some linguistic variations already captured in
the subjective database. 
Such common queries include ``clean room'', ``good breakfast'',
and ``nice location'' for the hotel domain and 
``tasty food'', ``friendly staff'', and ``ambience'' for the restaurant domain.
These phrases and their synonyms also frequently appeared in reviews.


Word2vec~\cite{word2vec} allows one to compute a vector representation
of a word or a short phrase (e.g. bi-gram). The vector representation is typically
a dense vector with hundreds of dimensions.
Two phrases have similar vector representations if they share similar
contexts in the text corpus, and so the two phrases are \emph{semantically} similar to each other.
The query predicates and the linguistic variations can contain
multiple words or short phrases. To compute their vector representations $\rep(\cdot)$,
we use the IDF-weighted sum method commonly used in the NLP community:
\vspace{-1mm}
\begin{equation}
\rep(p) = \sum_{w \in p} \wtv(w) \cdot \idf(w) .
\vspace{-1mm}
\end{equation}
Here, $p$ is the query predicate or the linguistic variation, 
$w$ is a word or short phrase of $p$, $\wtv(w)$ is the word vector of $w$,
and $\idf(w)$ is the Inverse Document Frequency (IDF) \cite{ir} of $w$ in the review corpus.
Intuitively, $\idf(w)$ measures the importance of the word $w$ so that less frequent words
are weighted higher in $\rep(p)$. For example, the short phrase ``very-clean'' has
a higher weight than ``clean'' since ``very-clean'' is less frequent than ``clean''.
Then, to measure the closeness of a query predicate $q$
to a linguistic variation $p$, we simply compute the cosine similarity of their representations:
\vspace{-1mm}
\begin{equation}
\similarity(q, p) = \cosine(\rep(q), \rep(p)) .
\vspace{-1mm}
\end{equation}

There are more sophisticated methods for computing short text representation (i.e., sentence embedding)
like Skip-Thought Vectors \cite{sent2vec1} and InferSent \cite{sent2vec2}. 
These methods have shown good performance in tasks like sentence classification, entailment, and 
similarity search. One can build an interpreter method that first converts the query predicate
into the sentence embedding then performs a similarity search over the linguistic domains.
However, these methods usually involve computation with a neural network
and similarity search which can be expensive. Such operations can lead to less efficient query processing.
On the other hand, due to its simplicity, the IDF weighted method 
enables \system\ to reduce the cost of these expensive operators with efficient indexing schemes.
We introduce one such method in Appendix B of \cite{opinedb_arxiv}.

The word2vec method can fail if there is no similar linguistic variation found in the database.
Specifically, when the highest similarity returned is below a certain threshold (e.g., 0.5),
\system\ will turn to the co-occurrence method, which we describe next.

\smallskip
 \textbf{Co-occurrence method:} 
We can decide whether or not a predicate $q$ should map 
to an expression $A.m$ according to
whether $q$ frequently co-occurs with linguistic variations of $m$ in the source text of the subjective database.
We use this method only when the query predicate
cannot be satisfactorily mapped to markers of the existing 
set of subjective attributes as described above.
For example, the predicate ``{\em is a romantic getaway}'' is not sufficiently similar to any linguistic variation (i.e., the highest similarity is below the threshold 0.5).
Hence, \system\ uses instead the co-occurrence method for this predicate. It discovers that this predicate
frequently occurs in positive reviews where ``{\em excellent service}'' and ``{\em five-star bathrooms}''
are also mentioned. Hence, it is likely that the query predicate is correlated to
the {\tt service} attribute and ``{\em exceptional service}'' is the closest
marker of that attribute. In addition, the query predicate is also
close to the {\tt style} attribute and ``{\em luxurious bathrooms}'' 
is the closest marker of that attribute to ``{\em five-star bathrooms}''. 

Specifically, given a query predicate $q$, \system\ first searches the source reviews 
of the subjective database to find all positive reviews where $q$ occurs. 
We measure the positiveness of a review by applying sentiment analysis~\cite{nltk}.
Among the set of related reviews, we find the top-$k$ reviews ranked 
by the following scoring function
\vspace{-1mm}
\begin{equation}
\mathsf{rank\_score}(d) = \mathsf{BM25}(d, q) \cdot \senti(d),
\end{equation}
where $d$ is a review text, 
$\mathsf{BM25}(d, q)$ is the classic Okapi BM25 \cite{ir} 
ranking function measuring relevance of $d$ and $q$ based on tf-idf and 
$\senti(d)$ is the sentiment score computed on the review $d$.
Efficient computation of the top-$k$ documents ranked by BM25 is well-studied
with mature implementations such as Elasticsearch \cite{elastic}.

Afterwards, we collect the set of linguistic variations 
extracted from the top-$k$ reviews. 
The most correlated attributes are
the ones with the highest
tf-idf score. Formally, for each subjective attribute $A$, 
we let $\freq_k(A)$ be the number of times that linguistic variations of attribute $A$ 
are extracted among the top-$k$ search result. 
Let $\idf(A)$ be the inverse document frequency of attribute $A$.
The final interpretation of a query predicate consists of a disjunction of $n$ expressions 
$A.m$ where (1) $A$ is an attribute with the top-$n$ highest $\freq_k(A) \cdot \idf(A)$
and (2) $m$ is the marker of $A$ with the highest frequency in the top-$k$ reviews. 
When the top-$n$ highest score is below a certain
threshold, \system\ considers the result to be less confident and will 
turn to the results of text retrieval. 

Table \ref{tab:cooccurrence} illustrates the strength of 
the co-occurrence method with outputs from real examples.

\setlength{\tabcolsep}{5pt}
\begin{table}[!ht]
\small
\centering
\vspace{-5mm}
\caption{\small{Example output of the co-occurrence method.}}\label{tab:cooccurrence}
\begin{tabular}{@{}p{1.5cm}@{}|l|l}
\toprule
& \textbf{Query Predicates}      & \textbf{Top-1 Interpretations}                     \\ \midrule
\textbf{Hotels}       
& for our anniversary            & staff.\emph{``helpful concierge''}      \\
& multiple eating options        & food.\emph{``good options''}        \\
& kid friendly hotel             & staff.\emph{``very kind staff''}   \\ \midrule
{\bf Restaurants}
& dinner with kids               & table.\emph{``high chair''} \\
& close to public transportation & general.\emph{``great place''}  \\
& private dinner                 & vibe.\emph{``quiet place''} \\ \bottomrule
\end{tabular}
\vspace{-2mm}
\end{table}






After interpretation, \system\ computes how well each resulting $A.m$
matches with a database entity by computing the degree of truth (Section \ref{sec:dot}).

\smallskip
{\bf Text-retrieval method:}
In the event that both word2vec and co-occurrence failed to 
interpret a query predicate with high confidence, 
we fall back to traditional information retrieval techniques to 
compute the degree of truth based on ranking scores of each entity w.r.t. the query phrase. 

Following a previous work \cite{opinion-based}, the text-retrieval method represents
each entity by a single document $D$ 
obtained by combining all source reviews of the entity.
Then for a subjective query predicate $q$, \system\ computes the ranking score 
simply as $\mathsf{BM25}(D, q)$. To convert the value into a degree of truth,
we set a constant threshold $c$ and apply the sigmoid function.
The returned degree of truth is computed as $\sigmoid(\mathsf{BM25}(D, q) - c)$.


\subsection{Computing the degrees of truth}
\label{sec:dot}

After the predicates have been interpreted into an expression over a set of $A.m$'s, 
\system\ now needs to compute how well the reviews of each entity represent each query predicate $q$. 
In other words, 
\system\ computes a degree of truth, which is a value between 0 (false) and 1 (true) 
for each interpreted predicate such as {\tt room\_cleanliness}.``very clean''. 

As mentioned in Section~\ref{sec:simple-predicates},  the degrees of truth for variations 
in the linguistic domain (i.e., $m = q$) of each subjective attribute can be pre-computed so that they can simply 
be looked up at query time.
For phrases that are outside the linguistic domain ($m \neq q$), the degrees of truth are computed during query time. 
These degrees of truth, once computed, can also be indexed and so they can be simply retrieved in future.

\system\ has access to the relevant marker summaries 
through the interpretations obtained. Next,
we describe how \system\ translates marker summaries into degrees of truth
w.r.t. a predicate.





\vspace{1mm}
\noindent
\textbf{Membership functions:}
\system\ constructs a {\em membership function}~\cite{klir1995fuzzy} to
compute the degree of truth of an interpretation $A.m$ based on the marker summaries of $A$ and the
interpreted marker $m$ with original query predicate $q$.
In effect, the membership function further aggregates the
marker summary to compute the degree of truth. For example, 
the marker summary [\text{``v. clean''}: 20, \text{``avg.''}: 10, \text{``dirty''}: 1, \text{``v. dirty''}: 0]
should have a value close to 1 (e.g., 0.95) for the query predicate ``really clean room''
since most reviews mentioned that the rooms are clean.
In contrast, the marker summary [\text{``avg.''}: 10, \text{``dirty''}: 10]
should have a much lower value (e.g., 0.2) for  ``really clean room'' since
half of the extraction results stated that the rooms are dirty.

\system\ uses machine learning to construct the membership functions.
Specifically, \system\ trains classification models from labeled tuples 
$\{(S_i, p_i, y_i)\}_{i \geq 0}$ where each $S_i$ is a marker summary, 
$p_i$ is a phrase and $y_i \in \{0, 1\}$ is a binary label that 
indicates whether or not $S_i$ satisfies $p_i$.
Binary classification is suitable for this task because
binary labels are less expensive to obtain compared to numeric labels.
Furthermore, many popular models such as \emph{Logistic Regression}
compute intermediate values that can be interpreted as a degree of truth in $[0, 1]$.
More specifically, logistic regression learns the binary classifier by first learning
a \emph{logistic loss function} which can compute the probability (so $[0, 1]$ is its range) 
of a label being positive given the input tuple.
As a result, we can directly use the probability output 
as the membership function by interpreting the probability as the degree of truth.

The model makes use of features constructed from precomputed information in each marker summary.
By doing so, \system\ can speed up query processing by avoiding scanning the full extraction tables.
Such features include the sizes of the markers, the average sentiment scores, and the centers of the phrase vectors
of the phrases mapped to each marker. \system\ trains high-quality models using these features
as the markers are expected to be good representations of the underlying linguistic domain.

In our experiment, we found
that with a set of 1,000 labeled tuples,
we obtained Logistic Regression classifier of 71\% to 75\% accuracy (Section \ref{sec:marker-exp})
on the hotel and the restaurant domains. This means that
the features constructed from the marker summaries are high-quality and hence,
the logistic loss is suitable as the membership function.
In addition, the use of the marker summaries in query processing results in a speedup up to 6.6x. 

\vspace{-1mm}
\section{Designing subjective databases}
\label{sec:aggregation}


The creation of the schema and of the data in a subjective database are closely 
intertwined. Next, we describe how \system\ (1) 
extracts opinions from text and (2) based on the extractions, it
constructs subjective attributes and marker summaries. Both processes are interactive in that the schema designer of \system\ provides input on 
what the important attributes are and what information needs to be extracted.

The problem of extracting opinions from reviews is a well studied problem in the NLP literature (e.g.,  \cite{ZhangWL18widm, liu2012sentiment, wang2017coupled, hamilton2016emnlp, wang2016recursive, vicente2014eacl, tai2013iiwas, qiu2011coling, hu2004aaai}). Our focus is not on developing new techniques for opinion mining, but rather devising techniques that enable the schema designer to quickly develop a good schema for the database.


\subsection{Extracting opinions from reviews} \label{sec:extractor}
\system\ extracts all the pairs of {\em aspect term} and associated {\em opinion term}.
For example, given the sentence:
\vspace{-1mm}
\begin{center}\emph{The room was very clean, but the staff was not so friendly . }\end{center}
\vspace{-1mm}
\system\ would extract pairs of the form
$$
\{(\text{``room''}, \text{``very clean''}), (\text{``staff''}, \text{``not so friendly''})\} . 
$$
Within each pair, the first element is the \emph{aspect term}, which represents
the target of the opinion. The second element is the \emph{opinion term}
containing an opinion on that aspect. This task is closely related to the
\emph{Aspect-Based Sentiment Analysis} (ABSA) problem \cite{absa, pontiki2014semeval, pontiki2015semeval}, 
which aims at finding the opinionated aspect terms from text and 
predicting their sentiment scores (i.e., positive or negative).
The solution was later extended to also extract the opinion terms
\cite{wang2017coupled, wang2016recursive}. Hence, these proposed techniques are
suitable for \system's extractor. 

\begin{figure}[!ht]
    \centering
    \includegraphics[width=0.48\textwidth, height=1.1cm]{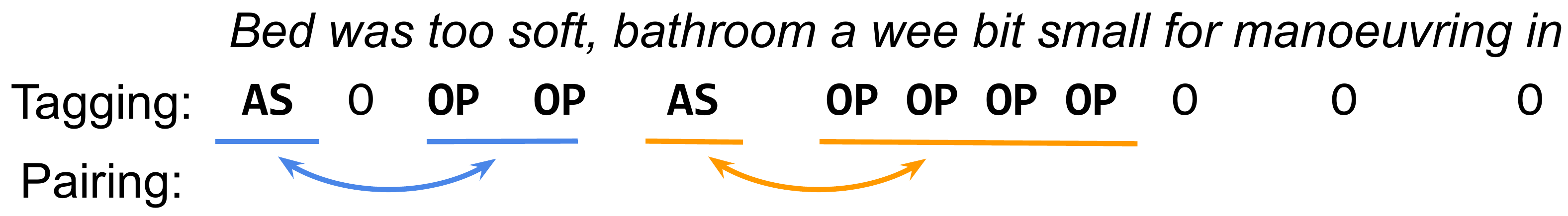}
    \vspace*{-7mm}
    \caption{\small{Tagging and pairing.}}
    \label{fig:extractor_example}
    \vspace*{-5mm}
\end{figure}

Following previous work, we design the extractor as a two-stage procedure: tagging and pairing.
This is
illustrated in Figure \ref{fig:extractor_example} with a real example of our extractor
applied to a hotel review.
During the tagging stage, the tokens of the input sentence are classified as (part of) an aspect term (AS), an opinion term (OP),
or irrelevant (O). In the pairing stage, the tagged aspect/opinion terms are paired to form
the extracted opinions. Here, we focus on optimizing
the quality of the tagging stage
since the pairing stage can be implemented
with a rule-based model and achieves comparable good performance to that of a learned model (More details in Appendix C of \cite{opinedb_arxiv}).

The reported results 
for electronics and restaurants reviews~\cite{wang2017coupled,wang2016recursive} were promising. However, 
the lack of labeled training data makes their trained 
deep learning models hard to generalize to other domains \cite{pontiki2015semeval}, (e.g., hotels).
Hence, instead of using these techniques, we built our extractor based on 
BERT~\cite{bert}, the recently developed
pre-trained NLP model that achieved start-of-the-art performance in major NLP tasks
including sentiment analysis and tagging. The transfer learning capability of BERT
allows the extractor model to first be trained on a large set of unlabeled text data, and then fine-tuned on a labeled training set of a much smaller size.
In our experiments we show that in the hotel domain, \system's extractor achieved a good 74.71\% 
F1 score when we use a pre-trained BERT model from \cite{bert} with 912 labeled review sentences for fine-tuning.
In contrast, the method based on \cite{wang2017coupled, wang2016recursive} achieves only 68.04\% F1 score. 
Labeling and training using these 912 review sentences data were done within a few hours.
Another advantage of using a pre-trained model like BERT is that
the model is fixed so that it does not require the schema designer to have NLP/ML expertise 
to program the neural networks or to tune the hyper-parameters.
As such, the whole process of developing the extractor is extremely efficient. 


\subsection{Designing the subjective attributes}
In the next step, the schema designer provides a set of subjective attributes
and \system\ maps each extracted pair to one of the attributes.
The design of the set of subjective attributes is analogous to the design of a relational database schema where we rely on the schema designer to decide what should or not be part of the schema. In future, we plan to provide the schema designer suggestions of possible subjective attributes by analyzing the extracted pairs.
In our experience, the number of subjective attributes is quite small (11 attributes for the restaurant domain
and 15 attributes for the hotel domain). 

We formulate the problem of assigning extracted pairs to attributes as a text classification problem. 
For example, \system\ would classify the above pairs to attributes as follows:
\begin{align*}
(\text{``room''}, \text{``very clean''}) & \longrightarrow \mathtt{room\_cleanliness} \\
(\text{``staff''}, \text{``not so friendly''}) & \longrightarrow \mathtt{staff}. 
\end{align*}

We need labeled data to train such a classifier.
To reduce the labeling cost, we construct the training set automatically via seed expansion.
For each attribute $A$, the designer provides a pair  $(E, P)$ of seeds where $E$ is a set of aspect terms
that $A$ describes and $P$ is a set of opinion terms that refer to those aspects.
For example, for the attribute $\mathtt{room\_cleanliness}$, the designer can provide:
\begin{align*}
E &= \{\text{``room'', ``bedroom'', ``carpet'', ``furniture''} \} \\
P &= \{\text{``clean'', ``dirty'', ``very clean'', ``very dirty'', ``stained'', ``dusty''} \}
\end{align*}
\system\ then  expands the seeds with synonyms using a word2vec model. 
The word2vec model is trained on the review corpus so it can capture similar phrases more accurately.
For example, phrases such as ``room'' may be expanded
into ``suite'', ``executive suite'', or ``apartment''. 
Next, for each each pair $(e, p)$ in the cross product of ($E \times P$) and attribute $A$,
\system\ constructs a labeled tuple $(\mathsf{concat}(e, p), A)$ where
$\mathsf{concat}(e, p)$ is the example and the attribute $A$ is the label.

This approach allows \system\ to train a high-quality attribute classifier with only little effort in creating the labeled dataset.
For example, with only 235 seeds of 15 restaurant attributes (expanded into a training set of 5,000 tuples),
\system\ is able to obtain a classifier with 88\% accuracy on the test set. 



\newcommand{\verbatimfont}[1]{\renewcommand{\verbatim@font}{\ttfamily#1}}

\subsubsection{Defining markers}\label{sec:domainmarkers}

Given the classification result, we can define the linguistic domain of each attribute
to be the set of all possible phrases (concatenations of the aspect and opinion terms) assigned to it. 
Next, the schema designer needs to specify a marker summary for each attribute and whether or not the attribute is linearly ordered.
\system\ alleviates the effort required for this step by providing two automated methods.
The design of these two methods is based on the observation that most linguistic domains can be modeled as one of the two types:
\vspace{-1mm}
\begin{itemize}\parskip=0pt
\item \textbf{Linearly-ordered domains. } 
The phrases of linguistic domains for attributes such as {\tt room\_cleanliness}
can be ordered linearly.
For example, 
$\text{``dirty''} < \text{``average''} < \text{``clean''} < \text{``very clean''}$.
In such cases, we can generate the markers by leveraging sentiment analysis \cite{nltk}.
More specifically, we sort the phrases by their sentiment scores $\senti(\cdot)$ and 
divide the linguistic domain equally into $k$ buckets.
The markers are designated as the linguistic variation
in the center of each bucket.
\vspace{-1mm}
\item \textbf{Categorical domains. } 
Linguistic domains can also be categorical, which means that the linguistic variations
can be categorized into a few topics. The {\tt bathroom} attribute is categorical -- the phrases
can be $\{\text{``luxurious'', ``modern'', ``old-styled''}\}$ where there is no clear linear order
but can be summarized into categories. In such cases, \system\ performs $k$-means clustering
on the linguistic domain. Afterwards,
\system\ suggests a set of markers by selecting the linguistic variations
that correspond to the centroid of each cluster.
\end{itemize}
\vspace{-2mm}

\subsubsection{Calculating marker summaries}
\label{sec:calculating-summaries}
Once the marker summary is defined, the next step is to aggregate the data from the reviews according to the markers. 
In general, the appropriate aggregation function depends on the semantics of the attribute.
Attributes like {\tt friendlyStaff} will change over time more frequently than the attribute 
{\tt quietLocation}. Hence, in the former case we may want to weigh recent reviews more heavily.
As another example, some aspects of a hotel (such as {\tt towelArt}) are mentioned much less frequently than others. 
In these cases, even a few mentions should be considered a strong signal. 

The aggregation function can also depend on the specific needs of the \system\ application.
For example, an application might decide to assign uniform weights to all reviews but another application
might want to assign higher weights to reviews marked as ``helpful'' by other users.
The full exploration of different possible aggregation functions is beyond the scope of this paper 
but we believe is an important aspect of building subjective databases. 

In the current implementation, \system\ aggregates phrases of reviews based on the number of occurrences in the reviews. 
For example, a \texttt{room\_cleanliness} marker summary is constructed for each hotel
by counting the number of phrases of reviews from the extraction 
relation that contain linguistic variations closest to ``very clean'', ``average'', ``dirty'', ``very dirty'', respectively for that hotel. 
Even though our model for linearly-ordered marker summaries allows for a phrase 
to contribute in part to different markers, our initial implementation
matches a phrase to only the best matching marker. We plan to explore techniques to weigh the proportion of contributions of each phrase to different markers in the future.
The resulting histogram is the marker summary for that hotel and is stored in the
{\tt room\_cleanliness} attribute of relation  HRoomCleanliness.

The marker summaries can be incrementally computed.
Furthermore, any result returned can be supported with evidence from the reviews on why that result is returned because \system\ keeps track of provenance of extracted phrases.





\newcommand{\sat}{\mathsf{sat}}
\newcommand{\quality}{\mathsf{quality}}
\newcommand{\satmax}{\mathsf{sat\text{-}max}}

\section{Experiments}\label{sec:experiments}
We  present our implementation of \system\ and
our experimental results.

\smallskip
\noindent
{\bf Overview~} Our first set of experiments investigates the need for experiential search. We show that a significant proportion of user requirements
are experiential in several domains.

In our second set of experiments, we compare the quality of \\ \system's query results with two baselines.
To do this, we constructed subjective databases for two domains: hotels and restaurants and designed a method for evaluating subjective query results. 
We show that even when \system\ 
is evaluated conservatively, \system\ outperforms the other two baselines 
over a variety of subjective queries. 

We also present experimental results on the quality of critical \system\ components. 
We show that the extractor which we use to produce the linguistic domains of 
our subjective database schema achieve F1 scores of close to 75\% for hotels and 85\% for restaurants
with only a small amount of labeled data provided. 
We also show that the marker summaries significantly accelerate subjective query processing (from 3.3x to 6.6x)
while maintaining the quality of the query results. 
Finally, we show that the predicate interpretation algorithm achieves a precision of up to 85\% 
on the combined use of word2vec and co-occurrence interpretation methods.

\subsection{The need for experiential search} \label{sec:subjective-queries}
Our very first experiment was to 
verify whether or not users search experientially.
To do this, we conducted a user study on Amazon Mechanical Turk \cite{mturk} to determine 
what are the important criteria users consider when
they search for certain types of entities.
More specifically, each MTurk worker is provided with a  question for a particular domain. For example, 
we posed the following task for hotels:
``{\em Suppose you have planned a vacation and are looking for a hotel. Other than cost, list 7 separate criteria you'd likely value the most when deciding on a hotel}''.

We asked 30 workers such questions for each domain.
We collected the answers and manually
(and conservatively) evaluate
whether each criterion is subjective or objective. For example, wifi is a criterion
that frequently shows up in hotel search but we interpret it as objective ({\em is there wifi}) rather than subjective ({\em fast and reliable wifi}). 
We asked similar survey questions for other domains such as 
Restaurant, Education, Career, Real Estate, Car, and Vacation. 
Table \ref{tab:subjective} summarizes
our results.
The table shows that a significant number of the desired properties are subjective for several domains. 
However, to the best of our knowledge, online services for these domains today provide keyword search over the reviews at best and 
do not directly support subjective querying.
\setlength{\tabcolsep}{5pt}
\begin{table}[!ht]
\centering
\vspace{-5mm}
\caption{\small{Subjective attributes in different domains.}}
\label{tab:subjective}
\small
\begin{tabular}{l|c|l}
\toprule
{\bf Domain}   & 
{\bf \%Subj. Attr} 
& {\bf Some examples} 
\\ \midrule
Hotel    & 69.0\% & cleanliness,  food, comfortable \\ 
Restaurant    & 64.3\% & food, ambiance, variety, service \\ 
Vacation & 82.6\% & weather, safety, culture, nightlife \\
College  & 77.4\% & 
dorm quality, faculty, diversity \\ 
Home     & 68.8\% & space, good schools, quiet, safe \\ 
Career   & 65.8\% &
work-life balance, colleagues, culture \\ 
Car      & 56.0\% & comfortable, safety, reliability  \\ \bottomrule
\end{tabular}
\vspace{-5mm}
\end{table}

\vspace{-1mm}
\subsection{Experimental settings} \label{sec:implementation}

Before we present the evaluation of \system, we describe how we measure the quality of our query results and how we generated subjective queries for our experiments.

\subsubsection{Implementation and experiment settings}
We implemented the extraction pipeline of \system\ in Python and used standard packages including 
Tensorflow \cite{abadi2016tensorflow} for neural networks,
NLTK \cite{nltk} for sentiment analysis, and Gensim \cite{gensim} for word2vec.
The core part of the pipeline is the adaptation of an existing neural network \cite{bert-ner}
based on BERT \cite{bert}\footnote{\small{We used the 12-layer uncased pre-trained model in all the experiments.}}, BiLSTM, and Conditional Random Field (CRF), adapted to opinion extractions. 

We implemented the querying engine of \system\ on top of {\sf PostgreSQL} \cite{postgres}.
We store the results of the extraction pipeline in a {\sf postgres} instance.
To execute a subjective SQL query, \system\ simply parses it with {\sf sqlparse} and applies
the query interpretation algorithms described in Section \ref{sec:life} to
translate the input query into an executable SQL query. 
The resulting SQL query computes the subjective predicates 
(translated into membership functions) as user-defined aggregates in {\sf postgres}.
For simplifying our experiments, we also implemented a version of \system\ without
using PostgreSQL.
Both implementations and all the experimental scripts are open-source and available in \cite{open-source}.

In our experiments, we designed two subjective database schemas 
(for hotels and restaurants) and their respective linguistic domains and marker summaries
with \system. The reviews for hotels and restaurants are from two real-world datasets: 
\url{Booking.com} dataset \cite{bookingdataset} with 515,739 reviews for 1,493 hotels
and a subset of the Yelp \cite{yelpdataset} dataset with 176,302 reviews for 860 restaurants in Toronto.
We trained neural networks using an AWS \textsf{p2.xlarge} server with one
NVIDIA Tesla K80 GPU. All other experiments were conducted on a server machine
with Intel(R) Xeon(R) CPU E5 2.10GHz CPUs.


\subsubsection{Generating subjective queries}  
\label{sec:experiments-query-predicates}
Since there is no benchmark for subjective queries, we had to create one. 
Along with the survey result in Table \ref{tab:subjective},
we also collected a set of subjective queries for the hotel and restaurant domains.
We collected 190 subjective query predicates for hotels and
185 query predicates for restaurants.
We construct conjunctions of query predicates by uniform sampling of the predicates. 
These conjunctions will form the {\sf where} clauses of our subjective SQL query.
We consider 3 sets of queries (easy, medium, and hard) for each domain. The number of conjuncts in easy, medium, and hard queries are 
2, 4, and 7 respectively. Each set consists of 100 subjective queries. 

We further increase the complexity of the queries by adding two variations 
to each query.
Specifically, each query is extended 
with each of the following options which are conditions over objective attributes.
For hotel queries, the two options are:
(1) find all hotels in London less than \$300 per night and
(2) find all hotels in Amsterdam.
For restaurant queries,
the two options are:
(1) find all low-priced restaurants (i.e., price range is one `\$' sign according to yelp) and
(2) find all Japanese restaurants.

Table \ref{tab:review_statistics}
shows some statistics of the hotels/restaurants under
each selection predicate.
For example,
there are 189 London hotels that are less than {\$}300 per night.  
The restaurant reviews
tend to be longer (higher average number of words) and 
more positive (as indicated by the average polarity 
returned by the NLTK sentiment analyzer).

\vspace{-6mm}
\setlength{\tabcolsep}{3.5pt}
\begin{table}[!ht]
\small
    \centering
    \caption{\small{Review statistics in booking.com and yelp.}}
    \label{tab:review_statistics}
    \begin{tabular}{c|c|c|c|c}
    \toprule
    & \#Entities & \#Reviews & avg \#words & avg polarity\\
    \midrule
    London,$<${\$}300 & 189 & 139,293 & 34.27 & 0.19 \\
    Amsterdam & 91 & 45,875 & 37.02 & 0.21 \\
    Low Price  & 112 & 22,302 & 104.01 & 0.71 \\
    JP Cuisine & 108 & 24,701 & 126.02 & 0.72 \\
    \bottomrule
    \end{tabular}
    \vspace{-3mm}
\end{table}

\setlength{\tabcolsep}{4pt}
\begin{table*}[!ht]
\centering
\small
	\caption{\small Query result quality by the top-10 results in the booking.com dataset (left) and the yelp dataset (right) where quality (NDCG@10) $=$ \#Satisfied-predicates / max. \#predicates can possibly be satisfied. Each entry has
    a confidence interval within $\pm$0.0168 (on 0.95 convidence level). }
	\label{tab:quality}
	\begin{tabular}{ccc}
		\begin{tabular}{c|ccc|ccc}
			\toprule
			\multirow{2}{*}{Method} &
			\multicolumn{3}{c}{London $\land$ <{}300} &
			\multicolumn{3}{c}{Amsterdam} \\
			& {easy} & {medium} & {hard} & {easy} & {medium} & {hard} \\
			\midrule
GZ12 (IR-based) & 0.75 & 0.75 & 0.76 & 0.66 & 0.68 & 0.71 \\ \midrule
ByPrice         & 0.65 & 0.67 & 0.68 & 0.37 & 0.38 & 0.39 \\
ByRating        & 0.62 & 0.64 & 0.65 & 0.51 & 0.53 & 0.54 \\
1-Attribute     & 0.71 & 0.72 & 0.72 & 0.70 & 0.71 & 0.72 \\
2-Attribute     & 0.76 & 0.77 & 0.78 & 0.71 & 0.72 & 0.73 \\ \midrule
OpineDB         & 0.80 & 0.82 & 0.84 & 0.81 & 0.83 & 0.84 \\
\bottomrule
		\end{tabular}
		
		&\ \ &
		
		\begin{tabular}{c|ccc|ccc}
			\toprule
			\multirow{2}{*}{Method} &
			\multicolumn{3}{c}{Low Price} &
			\multicolumn{3}{c}{JP Cuisine} \\
			& {easy} & {medium} & {hard} & {easy} & {medium} & {hard} \\
			\midrule
GZ12 (IR-based) & 0.68 & 0.71 & 0.73 & 0.72 & 0.75 & 0.78 \\ \midrule
ByPrice         & 0.56 & 0.60 & 0.62 & 0.64 & 0.67 & 0.69 \\
ByRating        & 0.65 & 0.70 & 0.72 & 0.70 & 0.73 & 0.75 \\
1-Attribute     & 0.74 & 0.77 & 0.78 & 0.76 & 0.78 & 0.80 \\
2-Attribute     & 0.77 & 0.78 & 0.79 & 0.80 & 0.81 & 0.83 \\ \midrule
OpineDB         & 0.76 & 0.80 & 0.82 & 0.78 & 0.81 & 0.84 \\ 
\bottomrule
		\end{tabular}

	\end{tabular}
\vspace{-4mm}
\end{table*}

\subsubsection{Evaluation metrics}

In the experiments, we use a metric based on the well-known 
Normalized Discounted Cumulative Gain (NDCG) \cite{ir} to measure 
how well the entities in the result satisfy the predicates in the subjective query.
More precisely, assume a subjective query $Q$ with $n$ query predicates \{$q_1, ..., q_n$\} 
returns top-$k$ entities $E$ = \{$e_1, ..., e_k\}$.
Let $\sat(q_i, e_j) \in \{0, 1\}$ denote whether $e_i$ satisfies $q_j$. 
The quality of the result is measured by counting the total number of query predicates 
that are satisfied by all the $k$ entities in the result:
\vspace{-2mm}
$$ \sat(Q,E) = \sum_{j=1}^k \left( \sum_{i=1}^n \sat(q_i, e_j) \right) / \log_2(j + 1) . $$
Intuitively, a higher $\sat(Q, E)$ indicates that the top-$k$ entities are more relevant to the 
searched query predicates in $Q$. The term $1/\log_2(j + 1)$ 
logarithmically penalizes the irrelevant entities closer to the top of the result.

\smallskip
\noindent
\textbf{Ground truth:} 
The ground truth of $\sat(q, e)$ is expensive to obtain as it requires one to 
go through all the reviews. So, we adopt a lighter-weight approach to generate 
the labels of $\sat(q, e)$.
First, we manually identify the subjective attribute $A$ in the schema closest 
to a query predicate $q$ (e.g., the closest attribute to ``with clean rooms''
is {\tt room\_cleanliness}). Afterwards, we ask a human labeler to label 
$\sat(q, e)$ by inspecting the marker summary of attribute $A$.
We also notice that many summaries have (1) only a few number of reviews,
(2) a large fraction of unmatched phrases, or (3) a large fraction of negative phrases. 
In these cases, we can further reduce the labeling cost
by avoiding human labelers altogether as labels 
can be automatically generated with high accuracy using a set of rules.
We verified 20 $\sat(q, e)$ labels for restaurants and 20 for hotels
by inspecting their source
reviews. Both sets have 19/20 labels well-supported
by the underlying reviews. 

To better illustrate the quality of the query results in our experiments, we also compute $\satmax(Q)$, the maximum score of the quality of a query result can be
when we know the ground truth. We have 
$\satmax(Q_i) = \max_{E} \{ \sat(Q_i, E) \} . $
For a set of queries $\{Q_1, \dots, Q_N\}$ with query results $\{E_1, \dots, E_N\}$, 
the \emph{quality of the workload} is then computed as
\vspace{-2mm}
$$ \quality(\{Q_1, \dots, Q_N\}) = \dfrac{1}{N}\sum_{i=1}^N \dfrac{\sat(Q_i, E_i)}{\satmax(Q_i)}. 
\vspace{-2mm}
$$

\subsection{Comparing \system\ with baselines}
\label{sec:experiments-baseline}
We compare \system\ with two baselines: 
(1) IR-based search engine (IR) and
(2) attribute-based query engine (AB).

The IR baseline is an implementation of \cite{opinion-based} (GZ12), which applied
the IR method Okapi BM25 \cite{ir} retrieval model to rank entities
based on the opinions they received. Following
\cite{opinion-based},
we also added the capability to perform query expansion and different methods for 
combining multiple query predicates to make the baseline more competitive.




The AB baseline represents what a user
can obtain through online services such as \url{booking.com} or \url{yelp.com} 
by freely trying combinations of queryable attributes to obtain the best results. Hence, it is a strong baseline for comparing \system.
For example, to search for a hotel, the user can rank the hotels
by price or rating or even the predefined filters for some subjective attributes. 
We scraped all 8 subjective attributes (Location, Cleanliness, Staff, 
Comfort, Facilities, Value for Money, Breakfast, Free Wifi) from booking.com. 
We assume that the user can choose two of the above attributes
and rank the hotels by their sums.
Among all the combinations of attributes, we pick the one that maximizes
the score $\sat(Q, E)$.

Similarly, for restaurant queries, the user can rank the restaurants by the
number of stars or by the total number of reviews received.
Additionally, the user can choose to filter the restaurants using one or two of
the 33 available categorical attributes in the dataset. 
Some examples are {\tt Attire, GoodForGroups, NoiseLevel,} and\\
{\tt Ambience}.
The combination with the maximal $\sat(Q, E)$ is picked.

Table \ref{tab:quality} shows the quality results on both datasets. 
Each column in the tables represents the type of query used. 
That is, whether it is easy, medium, or hard and extended with 
an objective predicate (e.g., in London). The first row tabulates the results for the IR method, followed by 4 variations of the AB baseline and finally, \system. 
As noted earlier, the numbers represent 
the proportion of the number of query predicates that are satisfied out of 
the maximal number of query predicates that can possibly be satisfied.

To ensure the results' statistical significance, we repeated the experiment on 10 different samples of query sets 
(i.e., in total 1,000 queries per setting)
and computed the averages with confidence intervals. The results are indeed statistically significant
as the maximal size of the confidence interval is no larger than $\pm$0.0168.

In both datasets, \system\ outperforms the IR baseline by a 
sizable margin
(by $\sim$0.05 to $\sim$0.15 for hotels queries and 
by $\sim$0.06 to $\sim$0.10 for restaurant queries). 
This is not surprising as the IR baseline retrieves
hotels with reviews that contain keywords in the query predicates (e.g., ``clean'')
even if the same reviews contain the opposite negative words (e.g., ``dirty'')
or may have used the phrase ``not clean''. 
On the other hand, \system's 
membership functions can carefully discern between entities based
on the frequencies of positive versus negative phrases. 
We show one such example in the Appendix D of the full version \cite{opinedb_arxiv}.



The AB baseline has similar performance with the IR baseline. 
The tables clearly show that the result quality increases when more subjective attributes are used.
The AB baseline also performs much better in the restaurant queries.
This is because the yelp datasets contain more queryable attributes than
the hotel dataset. These findings reaffirm our belief that utilizing subjective attributes 
is important for experience search engines. 
Still, \system\ outperforms the AB baseline especially when there are more subjective query predicates. 
We believe this is due to \system's ability to accurately map those query predicates to subjective attributes.

Observe that the result quality of \system\ is higher in the hotel domain than in the restaurant domain
resulting in larger margins of improvement compared to baselines. This is
because the hotel dataset contains many more reviews per hotel and thus the generated
marker summaries are more representative and statistically significant.
This result matches our intuition and suggests that \system\ 
brings more value to applications as the number of reviews grows.


\subsection{Quality of OpineDB components}

Next, we evaluate the quality of important parts of \system\ :
the extractor, the marker summaries, and the predicate interpreter.

\subsubsection{Extractor and subjective DB construction} \label{sec:extractor-exp}

We start by showing that \system's extraction module achieves the
state-of-the-art or better quality. 
Moreover, we show that with the recent advances in NLP,
we are able to achieve the good performance with only a small amount of training data.



We evaluate the extractor on 4 datasets summarized in Table \ref{tab:extractor-datasets}.
The first 3 datasets are from ABSA competitions:
SemEval 2014 Task 4 (Laptops and Restaurants) \cite{pontiki2014semeval} and 
SemEval 2015 Task 12 (Restaurants) \cite{pontiki2015semeval}. 
Each dataset contains a set of sentences labeled with \emph{aspect terms}
and \emph{opinion terms} corresponding to the opinion targets and detailed opinions mentioned 
in Section \ref{sec:extractor} respectively. 
The aspect term labels are from the original datasets
and the opinion term labels were added by \cite{wang2017coupled} and \cite{wang2016recursive}.
Since there is no existing labeled opinion extraction datasets for hotels,
we created our own (Booking.com Hotel) to train our extractor. 
The sizes of the datasets are listed in Table \ref{tab:extractor-datasets}.
Note that none of the datasets are big.
The extractors trained on the hotel dataset and the SemEval-14 restaurant dataset
are the ones used in the experiment reported in Table \ref{tab:quality}.

Similar to other extraction tasks like Named Entity Recognition (NER) \cite{conll},
the extraction quality is measured by the \emph{F1 scores} of the aspect terms and the opinion terms.
An aspect/opinion term is considered correctly extracted only when the
extracted term matches exactly with the ground truth term.
As shown in Table \ref{tab:extractor-datasets}, the model of \system's extractor ({\tt BERT+BiLSTM+CRF})
outperforms the previous state-of-the-art models in all the 4 datasets\footnote{\small{We collected the F1 scores of the SemEval datasets 
 from \cite{wang2016recursive,wang2017coupled} and retrained their model on the hotel dataset (10 times to get the average).}}.
The improvement ranged from $\sim$0.01\% to $\sim$6.67\%.
We noticed that the improvement is the highest for the hotel dataset
which has the least number of training sentences. We believe that this 
is because of the transfer learning ability of the BERT model as 
similar observations were also reported in \cite{bert} for cases with a small amount of training data.
We also found that the quality of extraction is robust to small training sets 
so that a high-quality extractor can be obtained at very low cost.
Specifically, for the hotel domain, we found in our experiment that
even with a training set of 20\% of the original size ($<$200 sentences), 
the F1 score remains close to 70\% which is still higher than the SOTA.

\setlength{\tabcolsep}{4pt}
\begin{table}[!ht]
    \small
    \centering
    \vspace*{-4mm}
    \caption{\small Datasets for training and evaluating the extractor of \system. 
    The last two columns list the combined F1 scores (averaging the aspect/opinion terms' F1 scores) 
    of the previous state-of-the-art (SOTA) models and the one \system\ used.
    Each score of our model is the average score of 10 models trained separately. The $\pm$ indicates the standard confidence interval 
    computed on a 0.95 confidence level. }
    \begin{tabular}{c|cc|c|cc} \toprule
 Datasets           & Train & Test & Total & SOTA   & Our Model  \\ \midrule
 SemEval-14 Restaurant & 3,041    & 800  & 3,841 & 85.52  & 85.53 $\pm$ 0.40  \\
 SemEval-14 Laptop     & 3,045    & 800  & 3,845 & 78.99 & 79.82 $\pm$ 0.35  \\
 SemEval-15 Restaurant & 1,315    & 685  & 2,000 & 72.21 & 75.40 $\pm$ 0.58  \\
 Booking.com Hotel    & 800      & 112  & 912   & 68.04  & 74.71 $\pm$ 0.72  \\ \bottomrule
\end{tabular}    
    \label{tab:extractor-datasets}
\vspace{-2mm}
\end{table}



\system\ also performs classification to map each pair of extracted aspect/opinion terms
into the set of subjective attributes. To train such classifiers for the hotel and the restaurant domain,
\system\ applies weak supervision with the seed expansion techniques described in Section \ref{sec:extractor}.
The schema designer provided 15 subjective attributes with 277 seed phrases for the hotel domain
and 11 attributes with 235 phrases for the restaurant domain. 
For each domain, the seed expansion generates a training set of 5,000 records
and we manually labeled 1,000 addition records for testing purpose.
Both classifiers performed well: the hotel attribute classifier achieves an accuracy of 86.63\%
and the accuracy of the restaurant attribute classifier is 88.29\%.

Overall, the process of creating the subjective DB is efficient.
As mentioned above, the effort of creating the extractor for Hotels was $<5$ hours of human labeling and \system's extractor performs better than SOTA techniques.
Writing each seed set took us no more than 2 hours with 1 developer. 
These costs are small compared to the entire process of developing a travel application.


\subsubsection{Marker summaries and membership functions} \label{sec:marker-exp}

In addition to being a key component of \system's data model, the 
 marker summaries benefit a subjective DB in two ways:
(1) accelerating query processing, and 
(2) creating high-quality features for entity ranking. In this section we experimentally evaluate these benefits.
For both the hotel and the restaurant domain,
we created 10 markers for each subjective attribute 
by applying the automatic approach described in Section \ref{sec:domainmarkers}.
For each set of queries (the London, Amsterdam, Low-Price, and JP Cuisine queries listed above),
we compared  \system\ with a small variant of it which does not leverage the markers.
Specifically, when the markers are used, the logistic regression (LR) model for membership scoring
uses features precomputed for each marker (see Section~\ref{sec:life} for details). 
Without the markers, the model uses another set of engineered features 
similar to the set when the markers are used with the addition of new features (e.g., the number/fraction
of phrases that are similar to the query predicate) directly computed from the extracted phrases. Each LR model is trained on 1,000 labeled
pairs of entity and query.

\vspace{-3mm}
\setlength{\tabcolsep}{3pt}
\begin{table}[!ht]
\small
  \centering
\caption{\small \system\ w.\ 10 markers (10-mkrs) vs. no marker (no-mkrs). The running time is per 100 queries.
For each query set, LR-accuracy is the test accuracy of the Logistic Regression model,
NDCG@10 measures the query result quality, and Runtime is the total running time of 100 queries in seconds.
Each value is the average of 10 repeated runs. The max.CI column contains the maximal confidence interval 
of each row (on a 0.95 confidence level).}
\label{tab:markers}
  \begin{tabular}{cc|cc|cc|c}
    \toprule
            & & London       & Amsterdam    & Low-Price    & JP Cuisine  & max.CI \\ \midrule
\multirow{3}{*}{\rotatebox[origin=c]{90}{10-mkrs}} & LR-accuracy & 0.71  & 0.75  & 0.73  & 0.73 & $\pm$0.016 \\
 & NDCG@10     & 0.82  & 0.83  & 0.79  & 0.81 & $\pm$0.012  \\
 & Runtime     & 18.84s & 9.89s  & 12.55s & 13.95s & $\pm$0.726 \\ \midrule
\multirow{3}{*}{\rotatebox[origin=c]{90}{No-mkrs}} & LR-accuracy & 0.71  & 0.76  & 0.71  & 0.71 & $\pm$0.02  \\
 & NDCG@10     & 0.76  & 0.83  & 0.81  & 0.83 & $\pm$0.016  \\
 & Runtime     & 68.66s & 33.00s & 70.05s & 92.68s & $\pm$4.689 \\ \midrule
 & Speedup & 3.65x	& 3.34x	& 5.59x &	6.65x & $\pm$0.237 \\ \bottomrule
  \end{tabular}
\vspace{-1mm}
\end{table}

Table \ref{tab:markers} summarizes the results.
There is a significant performance improvement when markers are used, ranging
from 3.34x (Amsterdam) to 6.65x (JP Cuisine).
The overall average time per query is 0.14 sec when markers are used and
0.93 sec without the use of markers. 
Note that this gap will be much larger in real-world review datasets
and queries over a larger number of entities.
We also observed
that the quality of the membership functions (LR-accuracy) and query results (NDCG@10) remain mostly unchanged even with the speedup in performance.
This is because on small training sets (1,000 in our case), 
a smaller number of good features can help improve accuracy without overfitting. 
By aggregating the extracted phrases onto the marker summaries, \system\ reduces the number of features
while keeping the most relevant information. 

\subsubsection{Query predicate interpretation}
\label{sec:rewriter-experiment}

We executed our predicate interpretation algorithms on the hotel and the restaurant sets of query predicates from 
Section \ref{sec:experiments-query-predicates}. 
For each subjective query predicate, we manually labeled it with the closest
subjective attribute that the predicate should be mapped.
An interpretation result is counted as correct if
the attribute matches exactly with the ground truth.

\setlength{\tabcolsep}{3.5pt}
\begin{table}[!ht]
\small
\centering
\vspace{-4mm}
\caption{\small{Accuracy of different methods for query interpretation with confidence intervals over 10 runs.}}
\label{tab:interpreter}
\begin{tabular}{c|c|ccc|c} \toprule
Query sets & size & w2v      &  co-occur & w2v+co-occur & max.CI \\ \midrule
Hotel queries & 190 & 84.05 & 72.63      & 84.89  & 0.60\\ 
Restaurant queries & 185 & 81.62 & 68.65 & 82.16  & 0.52 \\ \bottomrule
\end{tabular}
\vspace{-1mm}
\end{table}


Table \ref{tab:interpreter} shows the accuracy of the 
two methods (word2vec and co-occurrence) when used independently
in the predicate interpretation algorithm and when used in combination 
(with the fallback similarity threshold set to 0.8).
The word2vec method produces reasonably high-quality ($>$80\% accuracy) interpretations.
The co-occurrence method has relatively lower accuracy (68\% to 72\%),
but it still improves the accuracy of the base w2v method when combined 
(by 0.84\% for hotel queries and 0.54\% for restaurants).
This is because although the co-occurrence method is relatively less accurate, 
it captures nicely the hard cases (long and uncommon text) that 
the word2vec method fails to capture.

\section{Related work}\label{sec:related}


The fields of sentiment analysis and opinion mining~\cite{ZhangWL18widm, liu2012sentiment, wang2017coupled, hamilton2016emnlp, wang2016recursive, vicente2014eacl, tai2013iiwas, qiu2011coling, WilsonWH05naacl, hu2004aaai} have developed techniques for extracting subjective data from text. While sentiment analysis tries to decide whether a particular text is positive or negative about an object or an aspect of an object, opinion mining aims to summarize a large collection of sentiments in a way that is informative to the user. In contrast, \system\ incorporates subjective opinion data into a general data management system, and addresses the challenges involved in doing so.

As described, a primary challenge in \system\ is to answer 
subjective queries over opinion data. A subjective query can be complex involving multiple subjective attributes and objective attributes
in addition to filters that restrict the reviews of interest (e.g., 
prolific reviewers, or reviewers that agree with the user's taste).
To the best of our knowledge, 
\system\ is the first system to answer complex subjective queries over review data in a principled way.
Opinion-based entity ranking~\cite{opinion-based, makris2014webist} 
are the only works that considered subjective queries 
and utilized reviews for ranking entities. 
However, that work did not aggregate the reviews or support complex queries like \system. 
Trummer et al.~\cite{DBLP:conf/sigmod/TrummerHLSG15} note that many queries to web search engines are of subjective nature and consider the problem of aggregating subjective opinions about (entity, attribute) pairs (e.g., cute animals). Aroyo and Welty~\cite{DBLP:journals/aim/AroyoW15} note that in the process of annotating training data for machine learning, there are several fallacies in assuming that there is an objective truth for the annotations. They develop a measure that supports differing subjective opinions from annotators.
Finally, subjective databases are different from probabilistic databases~\cite{DBLP:series/synthesis/2011Suciu} in that the latter still assume that there is an objective ground truth but it is not known to the database. 

The second challenge \system\ faces is to enable  application designers to apply domain semantics to subjective data management.
We introduce the concept of marker summaries to provide the designer such flexibility. 
The designer can tailor the linguistic domains 
and also which distinctions in the data to highlight in marker summaries. 
For example, one can have a coarse-grained notion of bathroom cleanliness (clean vs. dirty) or finer distinctions (shower cleanliness, faucet etc.)
The extractor of \system\ for extracting opinion expressions and forming linguistic domains 
is closely related to opinion mining~\cite{liu2012sentiment, absa}.
The extraction task is known as aspect term extraction~\cite{hu2004aaai, hu2004kdd, brody10hltnaacl, yan13www, he17acl} 
and opinion lexicon construction~\cite{hu2004aaai, hu2004kdd, liu2012sentiment, tai2013iiwas, vicente2014eacl, qiu2011coling, liu2012sentiment, rothe2016naacl, fast2016chi, hamilton2016emnlp, chen2018coling} that are well-studied in opinion mining.  
Following the recent trend of applying deep learning to opinion mining,
\system\ leverages the BERT pre-trained model \cite{bert} and achieved quality surpassing
the state of the art while requiring a small amount of human labeling effort. 

\system\ explores a variant of fuzzy logic 
to combine the scores of multiple query predicates. Fuzzy logic has been used in a myriad of applications in AI, control theory, and even databases with the capability of reasoning with vague and/or partial predicates 
like ``warm'' or ``fast'' \cite{xin2018subjective, zadeh1996fuzzy,klir1995fuzzy,fagin1996combining}.
The efficient evaluation of fuzzy selection queries has been broadly studied in databases,
with the Threshold Algorithm \cite{fagin2003optimal} 
and its descendants \cite{ilyas2008survey} 
as the most widely used techniques.
In contrast to previous work where fuzzy logic is used to reason about  ``partial truth''
or subjective perception of objective attributes like temperatures or speed,  
our work considers processing queries on data that is itself subjective.

The problem of building natural language interfaces to databases is a long-standing one~\cite{NLI2DB} and more recent work (e.g.,~\cite{IyerKCKZ17, LiJ16, PopescuEK03, Poon13}) 
has focused on learning how to parse natural language into a corresponding semantic form (e.g., SQL) 
based on examples of pairs of such. 
\system\ does not translate natural language into SQL. Instead, it supports query predicates, which are short phrases, that are already embedded in an SQL-like query.
Furthermore, the main focus of these works is on parsing objective queries but 
\system\ interprets and evaluates subjective queries.

\section{Conclusion}\label{sec:conclusion}

As user-generated data becomes more prevalent, it plays a critical role when users make decisions about products and services. However, by nature, user-generated data touches upon subjective aspects of these services for which there is no ground truth.
We introduced subjective databases as a key enabling technology 
for supporting experiential search and built \system, a first such system. 
\system\ has also been used to power Voyageur, our experiential travel search engine~\cite{voyageur}. 
\system\ is based on a new data model that incorporates user-generated data into a database system that can support complex queries, but also gives the designer flexibility to tune the schema for the application needs. We described how \system\ processes queries that require semantic interpretation and demonstrated that \system\ outperforms alternative approaches.

Subjective databases introduce several new future research challenges. There are many improvements that can be made to how such a system interprets queries specified using natural language. Similarly, a subjective database system should be able to take into consideration a user profile to provide better search results in case the user chooses to share such a profile. In the longer run, the system should be able to suggest queries to the user based on their profile and based on what may be unusual in the domain. For example, if there are reviews claiming that an expensive hotel has dirty rooms, that would be important to point out to the user because it contradicts their expectations. More generally, the challenge is to model the user's expectations and point out the unexpected experiential aspects.
Finally, the topic of bias on the Web is a very timely one~\cite{Baeza-Yates18}, and review data clearly contains biases. One of the interesting areas for future research is to use the expressive query capabilities of a system like \system\ to uncover biases with the goal of helping users make better decisions about their purchases.

\smallskip
\noindent \textbf{Acknowledgement } 
We are grateful to the anonymous reviewers for their thorough reports and 
many suggestions that have greatly improved the paper. 
We also would like to thank Megagon Labs members 
Shuwei Chen, Sara Evensen, George Mihaila, John Morales, Natalie Nuno, and
Ekaterina Pavlovic for the great engineering work of developing \system.



\bibliographystyle{abbrv}
\bibliography{paper}

\appendix

\section{Fuzzy logic vs. hard constraints} \label{app:fuzzy}
\begin{figure}[!ht]
    \centering
    \includegraphics[width=0.35\textwidth]{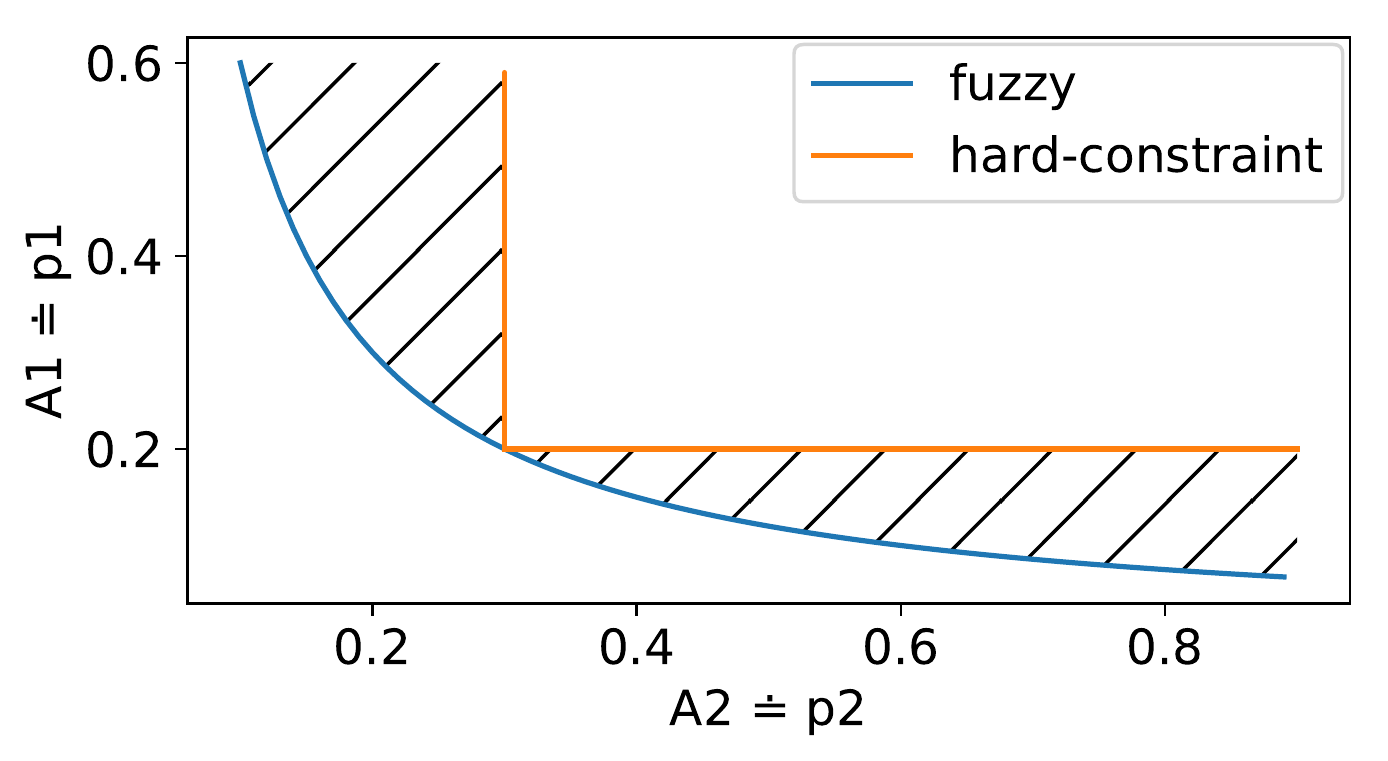}
    \vspace{-3mm}
    \caption{{\small Fuzzy constraints ($A_1 \doteq p_1$ $\otimes$  $A_2 \doteq p_2$) vs. hard constraints ($(A_1 \doteq p_1) > 0.2 \land (A_2 \doteq p_2) > 0.3$).}}
    \label{fig:fuzzy}
    \vspace{-3mm}
\end{figure}

We further compare the effect on the query results when interpreting
a subjective SQL query into fuzzy logic predicates vs. hard constraints.
As the number of conditions increases, the number of relevant entities that are potentially missed by the hard constraints only increases. 
Consider a conjunction of interpreted predicates ``$A_1 \doteq p_1$ $\otimes$  $A_2 \doteq p_2$'' where $\otimes$ is interpreted as multiplication.
The fuzzily combined degree of truth is represented by the blue curve 
(selecting entities with score at least 0.06) in Figure \ref{fig:fuzzy}.
The hard constraint $(A_1 \doteq p_1) > 0.2 \land (A_2 \doteq p_2) > 0.3$ is represented by the rectangular orange curve.
Clearly, the semantics under fuzzy logic (blue line) considers more entities than the other approach (orange line) and 
in particular, the blue line includes
those entities that fail to satisfy the hard constraints just by 
a little (the shaded area). 

As a consequence, the application or
end-user will need to manually
tune all the boundary parameters 
(e.g., 0.2 and 0.3 as in Figure \ref{fig:fuzzy})
to obtain a good set of results.
By interpreting the constraints with fuzzy logic, we naturally consider hotels that lie outside but close to the immediate boundaries.

\section{Indexing with the w2v-based sentence embedding} \label{app:w2v}
We present here one simple method for indexing with the {\sf w2v}-based method for sentence embedding.
We observe in our experiment that when the query $q$ is short, its most similar
variation $p$ typically differs from $q$ by at most 1 word, e.g., ``really clean room'' vs. ``very clean room''.
So for each word/bigram $w$ in the linguistic domain,
we precompute and index the word $w'$ closest to $w$ (i.e., with the minimal $|\wtv(w) \cdot \idf(w) - \wtv(w') \cdot \idf(w')|$).
At query time, we simply need to look up the index to try replace each word $w$ in $q$ with $w'$ and check whether the resulting
phrase $q'$ appears in the linguistic domain with a dictionary index. 
A full similarity search with a k-d tree index \cite{kdtree} is performed only when no $q'$ is found.
We found in our experiment that this simple index is very efficient:
it avoids performing the similarity search on 54.5\% of queries and results in
a 19.8\% speedup.

\section{Pairing models of the opinion extractor} \label{app:extractor}
In \system's extractor, the aspect and opinion terms are first extracted by the tagging model
then paired to form the set of extracted opinions. 
We considered two methods for pairing the aspect terms and the opinion terms.

The first method is an unsupervised rule-based method.
The intuition behind the rule-based method is that the linked aspect and opinion terms are usually
``close'' to each other. Furthermore, the distance between the terms can be
captured by their distance on the review sentence's parse tree. Thus, we can first compute the
parse tree of the review sentence and apply a greedy strategy to link the aspect/opinion term pairs
that are closest in the parse tree.

The second method is a supervised method based on sentence pair classification.
Each training example consists of a review sentence (e.g., ``the room was clean'')
and a phrase (e.g., ``clean room'') and the label is whether the phrase is a correct
extraction from the sentence. We constructed a training set of 1,000 sentence-phrase pairs (a mixture of postive and negative examples)
from the 912 hotel review sentence. We fine-tuned
a {\sf BERT} model and achieved an accuracy of 83.87\% on a test set of another 1,000 examples.

\section{OpineDB vs. the IR baseline} \label{app:example}
\begin{figure}[!ht]
    \centering
    \includegraphics[width=0.45\textwidth]{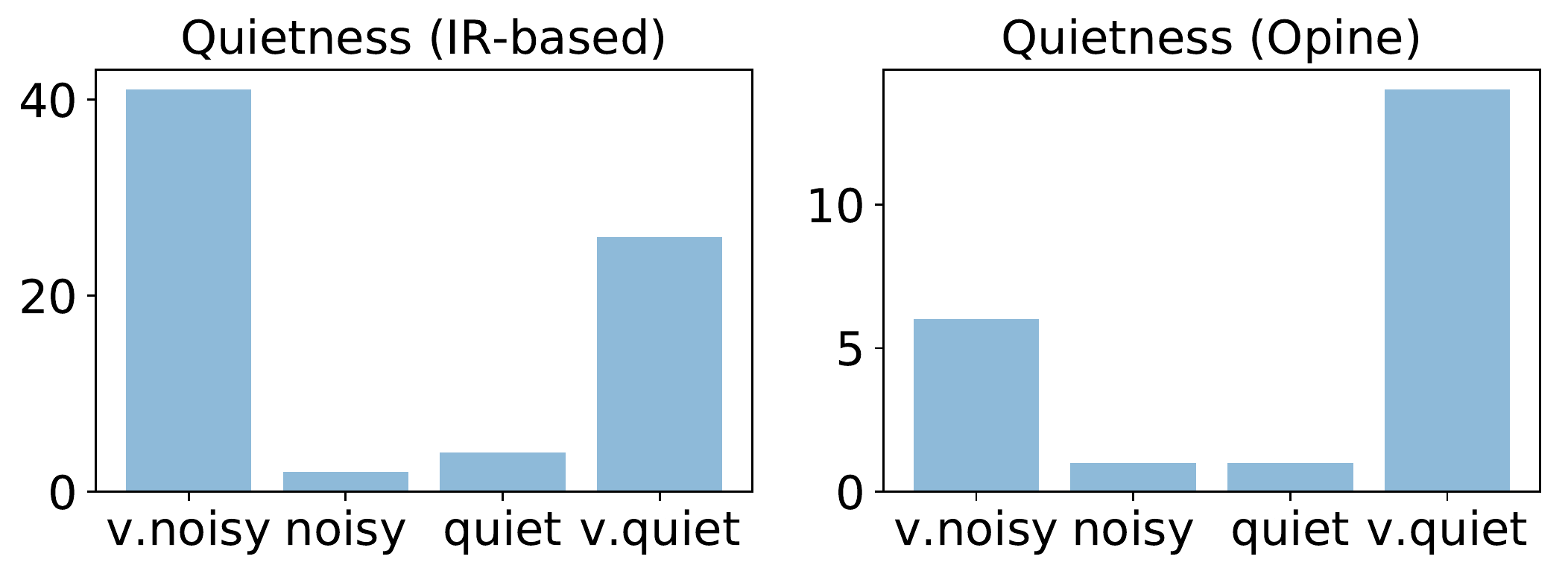}
    \vspace{-2mm}
    \caption{{\small The summaries of room\_quietness of two hotels returned by the IR baseline (left)
    and \system\ (right) for the query ``quiet room''.}}
    \label{fig:ir-example}
    \vspace{-1mm}
\end{figure}

We illustrate why \system\ is able to provide higher query result quality with an illustrative example.
Figure \ref{fig:ir-example} shows the marker summaries of two hotels returned by
\system\ and the IR baseline. This example shows that although
the result by the IR baseline can have high frequency of matched term with the query 
(``quiet room''), it can still contain negative opinions like ``very noisy room''
contradicting with the query. This issue is taken care of nicely by \system\ because of
its capability of aggregating the underlying phrases of the queried subjective attribute
{\sf room\_quietness}.

\end{document}